\documentclass[aps, prd, twocolumn, showpacs, superscriptaddress, groupedaddress]{revtex4} 
\usepackage{graphicx}    
\usepackage{amssymb}
\usepackage{dcolumn}
\usepackage{color}
\usepackage{subfigure, rotating, bm, array}
\usepackage[pagebackref=false, colorlinks=true]{hyperref}
\hypersetup{
linkcolor=blue,     
citecolor=blue,     
urlcolor=blue}   
\usepackage{soul}
\begin{document}
\title{On viability of isentropic perfect fluid collapse with a linear equation of state}
\author{Karim Mosani}
\email{kmosani2014@gmail.com}
\affiliation{BITS Pilani K.K. Birla Goa Campus, Sancoale, Goa-403726, India}
\author{Dipanjan Dey}
\email{dipanjandey.adm@charusat.edu.in}
\affiliation{International Center for Cosmology, Charusat University, Anand 388421, Gujarat, India}
\author{Pankaj S. Joshi}
\email{psjprovost@charusat.ac.in}
\affiliation{International Center for Cosmology, Charusat University, Anand 388421, Gujarat, India}

\date{\today}

\begin{abstract}
The gravitational collapse of a barotropic perfect fluid having the Equation of State (EoS) $p=k\rho$, where $k$ is constant, is 
studied here in the framework of general relativity. We examine the restrictions on the Misner-Sharp mass function, because of the introduction of such an EoS, in terms of the compatibility of a certain pair of quasi-linear partial differential equations,  obtained from Einstein's field equations. We find that except when this system of PDEs reduces to ODEs because of additional symmetries imposed on the spacetimes, or when they become compatible with each other in some special situations, consistent solution to perfect fluid collapse with linear EoS is not available. The end state of collapse with no such constraint of EoS has also been investigated.  Since considering arbitrary pressures in a collapsing cloud to study its end state is difficult as this requires information about the dynamics of collapse not known in  general, we consider small perturbations to mass profiles corresponding to inhomogeneous dust collapse. This, in turn, provides small pressure perturbations to the otherwise pressureless fluid. The dependence of visibility or otherwise of the singularity on initial conditions of collapse, in the presence or absence of such perturbations is studied numerically. As long as no linear EoS is imposed on the matter field, no incompatibility issue between its corresponding pair of PDEs arise, unlike the case when $k$ is restricted to be a
constant.
\\
\\
$\boldsymbol{key words}$: Black Hole, Cosmic Censorship Hypothesis, Gravitational Collapse, Naked Singularity.

\end{abstract}
\maketitle
\section{Introduction}
One of the most widely recognized problems in the foundation of gravitation physics is the Cosmic Censorship Hypothesis (CCH)\cite{penrose}. The hypothesis comes in two forms: the weak and the strong versions. The weak version of cosmic censorship hypothesizes that there can be no singularity observable from future null infinity, that is, the singularity is never visible to faraway observers in the universe.  The strong version suggests that singularities cannot be locally naked as well. By local nakedness we mean that a family of null geodesics can escape away from the singularity, however, it does not go beyond the boundary of the matter cloud but falls back to the singularity again
\cite{joshi}.

The singularity theorems 
\cite{hawking}, 
which prove the incompleteness property of nonspacelike geodesics for a wide class of spacetimes under physically reasonable conditions, thereby making the singularities inevitable in general relativity, have nothing to say about the properties of the singularity thus developed, and whether they are covered or not within a horizon of gravity. That is, no such inevitability rule applies to the existence 
of the event horizon that must cover the singularity. Thus the  
CCH remains unproven regardless of many grave efforts.   

In the homogeneous dust collapse case, all the matter shells collapse simultaneously to form an infinitely dense, infinitely curved
spacetime singularity. It was shown by Oppenheimer and Snyder \cite{oppenheimer}, and independently by Dutt \cite{datt} that this singularity is necessarily covered by an event horizon, thereby giving rise to a black hole as the collapse end state. However, when in-homogeneity is introduced in the density profile, it has been shown to give rise to both the possibilities, namely a Black Hole or Naked Singularity final states for the collapse \cite{joshi2,mena}. 
This is essentially because inhomogeneity affects the formation and geometry of the trapped surfaces, and thereby the apparent horizon, that develops as the gravitational collapse proceeds. If the time of formation of singularity precedes the time of formation of the apparent horizon, there is a possibility left for the outgoing null geodesics to just escape away from the central shell-focusing singularity. If this happens, then the strong cosmic censorship is violated, {\it i.e.} the singularity becomes at least locally naked.

Many collapsing models have been studied in which the strong CCH is necessarily violated \cite{dwivedi, deshinkar, goswami1}, however, whether or not these models are physically viable is still under discussion. For example, in the case of an inhomogeneous dust collapse, while it is physically more relevant compared to the homogeneous case, one could still argue that a physically viable scenario should have a non-zero pressure, and considering such pressured collapsing clouds might remove the possibility of the naked singularity thus formed. But it was seen later that  this is not true in general. Many collapsing models having non-zero pressures have been studied in detail wherein these are shown to give rise to naked singularities (see e.g. \cite{magli1, magli2, giambo1, harada1, harada2}, and \cite{joshi7} for a review).

Here, we are interested in a collapsing cloud satisfying a linear equation of state $p=k\rho$. While studying the analytic model of gravitational collapse of a spherically symmetric perfect fluid, one has a degree of freedom, hence a choice of specifying a function remains at disposal. However, the equation of state puts an additional constraint, thereby closing the system of Einstein's field equations. Gravitational collapse and final states for a barotropic spherical fluid were investigated by Giambo et al \cite{giambo}, in which no simplifying assumption of self-similarity (e.g. as used in \cite{ori1, ori2, ori3, carr, joshi4} was used. They showed that  
there exists a set of initial data giving rise to a naked singularity. However, whether or not this set has a non-zero measure, was unclear. The case of $\gamma$ law EoS and adiabatic condition was also studied \cite{harada}, wherein the result does not have the ansatz of self-similarity. The end state of  collapsing perfect fluid specifically with linear EoS has been investigated by Goswami and Joshi \cite{goswami}. 

Within such a perspective, further investigation carried out 
by us here reveals that the field equations have a collapsing solution only for those situations where two quasi-linear PDEs obtained from Einstein's field equations are compatible with each other.
For those mass functions, which have incompatibility issues amongst its corresponding two PDEs, it turns out that the linear EoS parameter $k$ cannot be constant anymore. This case increases the complications of solving Einstein's field equations. Hence an alternative tool at our disposal is to add a perturbation to the mass profile for dust collapse, to take an example. This, in turn, causes a non-zero pressure to come into the picture. We then follow the collapse formalism developed and used in \cite{joshi5}, to understand how the end state is affected due to this perturbation. In other words, we investigate whether or not the singularity formed is locally naked. 

It may be worth noting that the fascination about the end state of an unhindered gravitational collapse being visible lies in the fact that such a realm would be essentially governed by quantum gravity. One could speculate that if these effects can communicate to the far away observers, it could provide us with information about the quantum aspects of gravity, and this would help us possibly unify the fundamental forces of nature. 

The paper is organized as follows: Section II gives Einsteins field equations and the regularity conditions. Section III discusses the general solution of a particular type of partial differential equation, the quasi-linear PDE that comes from Einstein equations. 
This is then applied to get general solutions of PDE obtained by using $p=k\rho$, in section IV. Along with it, this section also discusses some examples such as the dust collapse, homogeneous perfect fluid collapse with non-zero pressure, and self-similar collapse,
all of which follow linear EoS with the corresponding PDE reducible 
to a simpler form. In Section V, a formalism for understanding the final state of collapse is discussed in as general way as possible. This formalism is then used to deal with different forms of perturbations to the mass profile of dust collapse in Section VI. The final section 
gives conclusions derived here.

\section{Einstein Equations}
The general metric representing gravitational collapse of a spherically symmetric cloud of perfect fluid, in comoving coordinates $t$ and $r$, having energy-momentum tensor $T^{\mu}_{\nu}=diag(-\rho,p,p,p)$, is given by 
\begin{equation} \label{cf1}
    ds^2=-e^{2\nu (t,r)}dt^2+e^{2\psi(t,r)}dr^2+R^2(t,r)d\Omega ^2.
\end{equation}
The Einstien's field equations with $8\pi G=c=1$ turn out to be 
\begin{eqnarray}
      \rho &=\frac{F'}{R^2R'},\label{efe1} \\
     p &=-\frac{\dot F}{R^2\dot R}, \label{efe2}\\
     \nu ' &=-\frac{p'}{\rho +p}, \label{efe3}\\
     2 \dot R' &= R'\frac{\dot G}{G}+ \dot R\frac{H'}{H},\label{efe4}
\end{eqnarray}
where,
\begin{equation} \label{cf6}
    G(t,r)=e^{-2\psi}R'^2; \hspace{0.5cm} H(t,r)=e^{-2 \nu}\dot R^2.
\end{equation}
The superscripts dot and prime denote here partial derivatives with respect to the coordinates $t$ and $r$ resectively. 
Here, $F$ is the Misner-Sharp mass function given by
\begin{equation}\label{msmf}
     F=R(1-G+H).
\end{equation}
$F$ denotes the mass of the cloud inside a shell of radius $r$ at time $t$. It can also be expressed as $F=r^3\mathcal{M}$ where $\mathcal{M}$ is a suitably differentiable function such that it does
not vanish 
or blow up in the limit of approach to the center. This is to preserve the condition of regularity of the Misner-Sharp mass at the regular center, and also to satisfy the condition that the energy density at the regular center, before the formation of central shell-focusing singularity, remains finite. An additional restriction is imposed on $\mathcal{M}$ as follows,
\begin{equation} \label{cusp}
    \mathcal{M}'(t,0)=0.
\end{equation}
This ensures that the energy density does not have any cusps at the regular center.

The metric component $R(t,r)$ is the physical radius of the cloud. To get a collapsing solution, the restriction  $\dot R<0$ needs to be imposed. We can express $R$ as follows:
\begin{equation}
    R(t,r)=r v(t,r),
\end{equation}
The scaling is done in such a way that $R(t_i,r)=r$, i.e. $v(t_i,r)=1$, where $t_i$ is the initial time for collapse. Now, the collapsing situation can be expressed as $\dot v<0$. The benefit of expressing $R$ in this manner is that the shell-focussing singularity formation can now be described by vanishing value of $v$. The physical radius $R$ can vanish when $r=0$ or $v=0$. However, the case $r=0$, $v \neq 0$ describes a regular center and cannot be termed as a singularity. Also, when $v\neq 0$, the energy density will not blow up at the regular center $r=0$. Hence by introducing the scaling function $v(t,r)$ in the way as above, the energy density blows up only at $v=0$, i.e. at the singularity. An alternate benefit of introducing the scaling function is that the collapse formalism could now be studied in the transformed $(r,v)$ coordinates, instead of $(t,r)$ coordinates.

It is observed that in the current situation, there is a total of five field equations in six unknowns, $\rho$, $p$, $\nu$, $\psi$, $R$ and $F$, i.e. two matter variables, three metric components, and the Misner-Sharp mass function. Hence, there is a freedom of choice of one free function. However, when we consider a barotropic fluid following a relation $p=p(\rho)$, then no freedom of choice remains
any longer. Our interest in this work lies in the linear relation $p=k\rho$, where $k$ is a constant and is called the linear EoS parameter. Using the first two field equations, the following partial differential equation is obtained in terms of  $\mathcal{M}$: 
\begin{equation}\label{ppde}
      k r{\mathcal M},r+((k+1)rv'+v){\mathcal M},v=-3k{\mathcal M}.
\end{equation}
Here, the superscript prime, as usual denotes a partial derivative with respect to $r$ in the $(t,r)$ coordinates. The subscripts $_{,r}$ and $_{,v}$ denote partial derivatives with respect to $r$ and $v$ respectively in $(r,v)$ coordinates.

\section{On the nature of quasi-linear PDE}
The first order quasi-linear PDE above of our interest is of the form 
\begin{equation}\label{gpde}
    M(x,y,z)z_{,x}+N(x,y,z)z_{,y}=P(x,y,z)
\end{equation}
This type of equation was first studied systematically by Lagrange, hence it is also referred to as Lagrange's equation 
\cite{dover}. 
The solution of Eq.(\ref{gpde}) is an integral surface which could be expressed implicitly as, 
\begin{equation}
    K(x,y,z)=z(x,y)-z=0.
\end{equation}
Eq.(\ref{gpde}) can hence be alternatively written as 
\begin{equation}
    (M,N,P). \nabla K=0,
\end{equation}
indicating that $(M,N,P)$, which is called the characteristic direction, lies in the tangent plane of the integral surface $z(x,y)$, provided 
that $\nabla K(x,y,z) \neq 0$.
The characteristic equations are given by 
\begin{equation}\label{gode}
    \frac{dx}{M}=\frac{dy}{N}=\frac{dz}{P}.
\end{equation}
If $a(x,y,z)=0$ and $b(x,y,z)=0$ are the solutions of the above characteristic equations, then the general solution of Eq.(\ref{gpde}) is obtained as, 
\begin{equation} \label{gsgpde}
    \mathcal{F}(a,b)=0,
\end{equation}
where $\mathcal{F}$ is an arbitrary function of $a$ and $b$. 
This is because of the following reason: 
\begin{enumerate}
    \item We differentiate $a(x,y,z)=0$ and $b(x,y,z)=0$ to get
    \begin{eqnarray}
     & a_xdx+a_ydy+a_zdz=0,\\ 
     &b_xdx+b_ydy+b_zdz=0.
    \end{eqnarray}
    \item Since $a=0$ and $b=0$ satisfy Eq.(\ref{gode}), we get
    \begin{eqnarray}
          & Ma_x+Na_y+Pa_z=0,\\
          & Mb_x+Nb_y+Pb_z=0.
    \end{eqnarray}
    Solving the above equations simultaneously leads to 
    \begin{equation} \label{ratiogpde}
        \frac{M}{\frac{\partial (a,b)}{\partial (y,z)}}= \frac{N}{\frac{\partial (a,b)}{\partial (z,x)}}= \frac{P}{\frac{\partial (a,b)}{\partial (x,y)}}.
    \end{equation}
    where the denominators are the Jacobian matrices.
    
\item Now, differentiating Eq.(\ref{gsgpde}) with $x$ and $y$ gives respectively,
\begin{equation}
\mathcal{F}_{,a}(a_{,x}+a_{,z}z_{,x})+ \mathcal{F}_{,b}(b_{,x}+b_{,z}z_{,x})=0,
\end{equation}
and
\begin{equation}
\mathcal{F}_{,a}(a_{,y}+a_{,z}z_{,y})+ \mathcal{F}_{,b}(b_{,y}+b_{,z}z_{,y})=0.
\end{equation}

Solving these two equations and using Eq.(\ref{ratiogpde}) gives Eq.(\ref{gpde}), indicating that Eq.(\ref{gsgpde}) is indeed a general solution of the partial differential Eq.(\ref{gpde}).
\end{enumerate}

Now, we discuss  particular solution of Eq. (\ref{gpde}). For this, an initial curve through which an integral surface $z(x,y)$  passes, has to be specified. The condition for existence and uniqueness of such a solution is materialized in what is known as the Cauchy problem. This problem can be formulated in eight different ways, all of which are equivalent to one another \cite{dover, bernstein}. One of the representation, and a particular case of our interest, stated geometrically is as follows:

\textit{Cauchy Problem}: For a given curve $\mathcal{T}$, $\exists$ a surface $z=\phi (x,y)$ passing through $\mathcal{T}$ such that at every point on the surface,  the direction of the normal $\nabla K=(z_x,z_y,-1)$ is such that Eq (\ref{gpde}) holds.

We refer to \cite{dover} for further details on Cauchy problem, 
and a classical theorem given by Kowalewski.

\section{The compatibility issue with constant $k$}
We now apply the techniques mentioned above to solve Eq(\ref{ppde}). 
The parametric form of the integral surface which is a solution of Eq.(\ref{ppde}), is expressed as 
\begin{equation}
    r=r(s_1,t_1), \hspace{0.5cm} v=v(s_1,t_1), \hspace{0.5cm} \mathcal{M}=\mathcal{M}(s_1,t_1).
\end{equation}
where $s_1$ and $t_1$ are parameters of the integral surface $\mathcal{M}(r,v)$ satisfying Eq.(\ref{ppde}). The corresponding characteristic equations are
\begin{eqnarray}
     &\frac{dr}{dt_1}  =kr, \label{ch11} \\
     &[k+1]r\frac{dv}{dr}-\frac{dv}{dt_1}  =-v, \label{spde} \\  
     &\frac{d\mathcal{M}}{dt_1}  =-3k\mathcal{M}.\label{ch13}
\end{eqnarray}
To find general solution of Eq.(\ref{ppde}), we solve Eq.(\ref{ch11}) and Eq.(\ref{spde}) simultaneously to get
\begin{equation}
     a(r,v,\mathcal{M})=rv=c_1.
\end{equation}
Similarly, solving Eq.(\ref{ch11}) and Eq.(\ref{ch13}) simultaneously gives
\begin{equation}
    b(r,v,\mathcal{M})=r^3\mathcal{M}=c_2.
\end{equation}
Here, $c_1$ and $c_2$ are constants. Hence, Eq. (\ref{gsgpde}) tells us that the general solution of Eq.(\ref{ppde}) is
\begin{equation}\label{gsppde}
    \mathcal{F}(R,F)=0.
\end{equation}

Now, we try to get a particular class of solutions satisfying Eq.(\ref{ppde}) by imposing suitable initial conditions representing an initial curve. Out of the three characteristic equations, Eq.(\ref{spde}) is itself a quasi-linear PDE. The characteristic equations for this PDE are as follows:
\begin{eqnarray}
\frac{dr}{dt_2} &=r(k+1) ,\label{ch21} \\
\frac{dt_1}{dt_2} &=-1 , \label{ch22}\\
\frac{dv}{dt_2} &=-v \label{ch23},
\end{eqnarray}
where the integral surface, which is the solution of Eq.(\ref{spde}) can be expressed in parametric form as 
\begin{equation}
    r=r(s_2,t_2), \hspace{0.5cm} t_1=t_1(s_2,t_2), \hspace{0.5cm} v=v(s_2,t_2).
\end{equation}
To find the general solution satisfying Eq.(\ref{spde}), we solve Eq.(\ref{ch21}) with Eq.(\ref{ch22}) and Eq.(\ref{ch22}) with Eq.(\ref{ch23}) simultaneously to get
\begin{equation}
    r^{\frac{1}{k+1}}e^{t_1}=c_3
\end{equation}
and
\begin{equation}
    ve^{-t_1}=c_4
\end{equation}
respectively, provided $k\neq -1$. This gives the general solution of Eq.(\ref{spde}) as 
\begin{equation} \label{gsspde}
    \mathcal{G}\left (r^{\frac{1}{k+1}}e^{t_1},  ve^{-t_1}  \right )=0.
\end{equation}
Characteristic curves for the characteristic equations Eq.(\ref{ch21}), Eq.(\ref{ch22}) and Eq.(\ref{ch23}) in terms of the parameters $s_2$ and $t_2$ are follows:
\begin{eqnarray}
    r(s_2,t_2) &=r_0(s_2)e^{(k+1)t_2}, \\ 
    t_1(s_2,t_2) &=-t_2+t_{{1}_0}(s_2), \\ 
    v(s_2,t_2) &=v_0(s_2)e^{-t_2}.
\end{eqnarray}
where the subscript `$0$' denotes the value of the function corresponding to $t_2=0$, for e.g. $r_0(s_2)=r(s_2,0)$. The solution of Eq.(\ref{spde}) depends on our choice of initial conditions.
Coming to the characteristic equations of Eq.(\ref{ppde}), the characteristic curves corresponding to Eq.(\ref{ch11}) and Eq.(\ref{ch13}) are obtained as 
\begin{eqnarray}
    r(s_1,t_1) &= r_0(s_1)e^{kt_1},\label{sch11} \\
    \mathcal{M}(s_1,t_1) &= \mathcal{M}_0(s_1)e^{-3kt_1}, \label{sch13}
\end{eqnarray}
where the subscript `$0$' denotes the value of the function corresponding to $t_1=0$. 
At the initial surface $v=1$ we have,
\begin{equation} \label{icppde}
   \mathcal{M}(r,1)=f(r)
\end{equation}
where $f(r)$ is a suitably differentiable function. The initial curve of the integral surface $\mathcal{M}(r,v)$ then is represented in parametric form as
\begin{equation}
    r_0(s_1)=s_1, \hspace{0.5cm} v_0(s_1)=1, \hspace{0.5cm} \mathcal{M}_0(s_1)=f(s_1).
\end{equation}
Using this, along with Eq.(\ref{sch11}) and Eq.(\ref{sch13}) gives
\begin{equation}\label{mrt1}
    \mathcal{M}(r,t_1)=f(re^{-kt_1})e^{-3kt_1}.
\end{equation}
Multiplying both sides of above equation by $r^3$ gives
\begin{equation}\label{Fequalsf1}
    F-f_1(re^{-kt_1})=0,
\end{equation}
where $f_1(x)=x^3f(x)$. In order for Eq.(\ref{Fequalsf1}) to be a particular solution of Eq.(\ref{gsppde}), $r$ should be non-minimally coupled with $v$ everywhere so that $F-f_1(R)=0$. This  is required because a particular solution of Eq (\ref{gsppde}) should be in terms of only $F$ and $R$. From this argument, we obtain
\begin{equation} \label{vt1}
   ve^{kt_1}-1=0. 
\end{equation}

However, this can never be a particular solution of Eq.(\ref{gsspde}) for the given initial condition except when $k=-1$. 
This is because, the above equation cannot be produced from any combination of the terms $r^{\frac{1}{k+1}}e^{t_1}$ and  $ve^{-t_1}$ for $k\neq1$.  In other words, Eq.(\ref{vt1}) does not satisfy Eq.(\ref{spde}). Hence, we conclude that Eq.(\ref{gsppde}) and Eq.(\ref{gsspde}) are not compatible with each other in general. However, there can be exceptions when the Eq.(\ref{ppde}) and Eq.(\ref{spde}) are compatible with each other. This happens in some cases when Eq.(\ref{ppde}) becomes an ODE, or under some additional symmetries.  Some examples of this type, having constant $k$ are as follows:

\begin{enumerate}
    \item When $k=0$, the Eq.(\ref{ppde}) reduces to 
    \begin{equation}
        (rv'+v)\mathcal{M}_{,v}=0
    \end{equation}
    which reduces to $\mathcal{M}_{,v}=0$, since we assume no crossing of the shell by imposing the condition $R'>0$, thus implying $\mathcal{M}$ is a function of $r$ only. It is observed that in case of dust, the general solutions Eq.(\ref{gsppde}) and Eq.(\ref{gsspde}) do not arise at all and hence incompatibility issue of these equations is not faced.
   
    \item For homogeneous collapse with non-zero pressure, we get $v=v(t)$, and hence the partial differential Eq.(\ref{ppde}) reduces to 
    \begin{equation}
           k r{\mathcal M},r+v{\mathcal M},v=-3k{\mathcal M},
    \end{equation}
    solving which, the general solution is obtained as 
    \begin{equation}
        \mathcal{H} (r^{\frac{1}{k}}v^{-1},\mathcal{M}v^{3k}  ).
    \end{equation}
    Here, there is no second PDE like Eq.(\ref{spde}) and hence no question of compatibility arises.
   
    \item In case of self-similar gravitational collapse  \cite{ori1, ori2, ori3, carr, joshi4}, $\nu$, $\psi$ and $v$ can be expressed as  functions of $X=t/r$, where $X$ is called the similarity parameter. The  function $\mathcal{M}$ is found using Eq.(\ref{msmf}) as
    \begin{equation}
        \mathcal{M} =\frac{\zeta (X)}{r^2},
    \end{equation}
    where 
    \begin{equation}
        \zeta (X)=v(1+e^{-2\psi}(Xv_{,X}-v)+ e^{-2\nu}(v_{,X})^{2}).
    \end{equation}
    This reduces the partial differential Eq.(\ref{ppde}) to an ordinary differential equation in $X$ as follows:
    \begin{eqnarray}
        k\zeta v_{,X}-(k+1)Xv_{,X}\zeta _{,X}+     v\zeta _{,X}=0.
    \end{eqnarray}
    Since this is an ODE, we will not have a situation in which the general solutions Eq.(\ref{gsppde}) and Eq.(\ref{gsspde}) arise and hence the compatibility issue is not required to be tackled.  
\end{enumerate}

What the above considerations show is, except when the above system of PDEs reduce to ODEs due to additional symmetries imposed on the spacetime, such as e.g. self-similarity, or in some rather special situations when these become compatible with each other, consistent solutions to perfect fluid collapse with a linear equation of state $p=k\rho$ will not be available. In a sense, this explains the severe lack of such solutions to the Einstein equations in general relativity in the literature so far. 
 
 When $k$ is not restricted to being a constant, the general solution of Eq. (\ref{ppde}) is same as Eq.(\ref{gsppde}) and that of Eq.(\ref{spde}) is 
 \begin{equation} \label{gsspde2}
     \mathcal{J}\left (t_1+\int\frac{dr}{(k+1)r}, ve^{-t_1}\right )=0.
 \end{equation}
 A suitable initial condition, Eq.(\ref{icppde}) gives
 \begin{equation}
     F=f_1 \left ( r e^{-\int kdt_1} \right )
 \end{equation}
Analogous to the way it was done in the case of constant $k$ before, for this to satisfy Eq.(\ref{gsppde}), $r$ should be non-minimally coupled with $v$ everywhere, from which
\begin{equation} \label{vt12}
   e^{-\int k dt_1}=v 
\end{equation}
is obtained. Differentiating this with respect to $t_1$ gives
\begin{equation}
    \frac{dv}{dt_1}=-vk.
\end{equation}
The above equation along with Eq.(\ref{spde}) gives the following expression of $v$:
\begin{equation} \label{vct1}
    v=\frac{q(t_1)}{r},
\end{equation}
where $q(t_1)$ is an arbitrary suitably differentiable function. Solving Eq.(\ref{vt12}) with Eq.(\ref{vct1})  gives the expression of $k$ as
\begin{equation}
    k(t_1,r)=\frac{q\frac{dr}{dt_1}-r\frac{dq}{dt_1}}{r^2v(t_1,r)}.
\end{equation}
Hence, there is no incompatibility issue in this case even if the partial differential Eq.(\ref{ppde}) is not reducible to any simpler form, unlike the above three cases.  

The next important problem which needs to be addressed is a formalism to determine the end state of a collapse of a matter cloud with no incompatibility issue such as above. This is presented in the following section.

\section{Final state of collapse}
 We now recall the formalism developed earlier \cite{joshi5, joshi6} to study the end state of collapse for the following cases:

 \begin{itemize}
     \item Collapsing clouds with linear EoS, having no incompatibility issue amongst the two PDEs, Eq.(\ref{ppde}) and Eq.(\ref{spde}). The three configurations, namely dust, homogeneous perfect fluid, and self-similarity, mentioned in the previous section, come under this category. These models are very well understood.
     \item Collapsing clouds not following linear EoS, however, where the corresponding partial differential Eq.(\ref{ppde}) may not be reducible. This category includes cloud having arbitrary pressures.
 \end{itemize}
 

First, we define a suitably differentiable function $A(r,v)$ as follows:
\begin{equation}\label{av}
    A_{,v}=\nu ' \frac{r}{R'}.
\end{equation}
Integrating Eq.(\ref{efe3}) along with using the above equation allows us to express $G$ as
\begin{equation} \label{G2}
    G(r,v)=b(r)e^{2A(r,v)},
\end{equation}
where $b(r)$ is an integration constant (with respect to $v$), and is related to the velocity with which the matter shells fall in. Near the regular center, $b$ can be expressed as
\begin{equation}
    b(r)=1+r^2b_0(r).
\end{equation}
Now, it is easy to make an analogy of $b_0(r)$ with that of the Lemaitre-Tolman-Bondi (LTB)  model \cite{lemaitre, tolman, bondi}, which has the criterion that depending on the polarity of $b_0$, we can get bound, unbound and marginally bound LTB collapse for $b_0<0$, $b_0>0$ and $b_0=0$ respectively. Equation of motion can be found using Eq.(\ref{msmf}) as 
\begin{equation} \label{eom}
    \dot v=-e^{\nu} \sqrt{\frac{\mathcal{M}}{v}+\frac{be^{2A}-1}{r^2}}.
\end{equation}
After substituting for pressure and density in terms of Misner-Sharp mass function $F$ and expressing it as $F=r^3\mathcal{M}$, we get
\begin{equation}\label{nudash}
    \nu '=\frac{\mathcal{M}_{,vr}v+(\mathcal{M}_{,vv}v-2\mathcal{M}_{,v})v'}{(3\mathcal{M}+r\mathcal{M}_{,r}-\mathcal{M}_{,v})v}R',
\end{equation}
where $v'$ is the derivative of $v$ with respect to $r$ in $(t,r)$ coordinates and the subscript $_{,r}$ denotes the derivative with respect to $r$ is $(r,v)$ coordinates.  For small $r$, $\mathcal{M}$ can be expanded as a Taylor series around the regular center $r=0$ as 
\begin{equation} \label{mf}
    \mathcal{M}=M_0+M_2 r^2.
\end{equation}
Here $M_1$ is taken as zero to avoid cusp at the center in the density profile as in Eq.(\ref{cusp}). $A(r,v)$ near the regular centre can be expanded as $A(r,v)=A_0(v)+A_1(v)r+A_2(v)r^2+....$. However, the form of mass function which  we consider leads to $A_0=A_1=A_3=0$. Using Eq.(\ref{av}) along with the expression for $\nu'$, i.e. Eq.(\ref{nudash}),  allows us to write $A(r,v)=A_2(v)r^2$ where
\begin{equation}\label{A2int}
    A_2(v)=\int_{v}^{1}\frac{2M_{2,v}+\bigg (M_{0,vv}-\frac{2M_{0,v}}{v} \bigg)v' _{,r}}{3M_0-M_{0,v}v}dv.
\end{equation}
Eq.(\ref{eom}) can be integrated to get 
\begin{equation}\label{singcurve}
    t(r,v)=t(r,1)+\int _{v}^{1}\frac{e^{-\nu}}{\sqrt{\frac{M}{v}+\frac{be^{2A}-1}{r^2}}}dv.
\end{equation}
Here
$t(r,v)$ is a $C^2$ function at least, in order for the regularity condition to hold, and hence can be expanded in the neighbourhood of the regular center as
\begin{equation}\label{singcurve2}
    t(r,v)=t(0,v)+\chi _1(v)r+\chi _2(v)r^2+O(r^3),
\end{equation}
where $\chi _1(v)=\frac{dt(r,v)}{dr}|_{r=0}$ and $\chi _2(v)=\frac{1}{2}\frac{d^2t(r,v)}{dr^2}|_{r=0}$ at the regular center $r=0$.

Now, the equation for outgoing radial null geodesics is given by 
\begin{equation}\label{efong}
    \frac{dt}{dr}=e^{\psi-\nu}.
\end{equation}
For the geodesics to cease at the singularity in the past, we should have $R \to 0$ as $t \to t_s$ along these curves. Eq. (\ref{efong}) in terms of $R$ and $u=r^{\alpha}$, where $\alpha>1$,  can be expressed as
\begin{equation}
    \frac{dR}{du}=\frac{1}{\alpha}\frac{R'}{r^{\alpha-1}}\left (1+ \frac{\dot R}{R'}e^{\psi-\nu} \right ).
\end{equation}
Using Eq.(\ref{msmf}) and some re-arrangement gives us
\begin{equation} \label{dRbydu}
    \frac{dR}{du}=\frac{1}{\alpha} \left ( \frac{R}{u}+ \frac{\sqrt{v}v'r^{\frac{5-3\alpha}{2}}}{\sqrt{\frac{R}{u}}} \right ) \left (\frac{1-\frac{F}{R}}{\sqrt{G}(\sqrt{G}+\sqrt{H})} \right ).
\end{equation}
For a singularity to be naked (at least locally), the tangent to the future directed radial null geodesics, which cease at the singularity in the past, we should have $\frac{dR}{du}>0$ at the singularity in the $(R,u)$ plane \cite{joshi6}, and also  it should be finite. Using L'Hospital's rule, we get,
\begin{equation} \label{x0}
    X_0= \lim_{(R,u)\to (0,0)} \frac{R}{u}=\frac{dR}{du}.
\end{equation}
Along any constant $v$ surface, $dv=v'dr+\dot v dt=0$, and hence, $\sqrt{v}v'$ could be obtained from Eq.(\ref{eom}), which is then substituted in  Eq.(\ref{dRbydu}) along with using Eq.(\ref{x0}) to obtain,
\begin{equation} \label{x01}
    X_0^{3/2}=\frac{1}{\alpha-1}\sqrt{M_0(0)}\chi_1(0)r^{\frac{5-3\alpha}{2}}
\end{equation}
if $\chi_1 \neq 0$, and
\begin{equation} \label{x02}
     X_0^{3/2}=\frac{1}{\alpha-1}\sqrt{M_0(0)}\chi_2(0)r^{\frac{7-3\alpha}{2}}
\end{equation}

if $\chi_1=0$ and $\chi_2 \neq 0$. Here, since $r \to 0$ in the above two equations, it could be concluded that for a certain value of $\alpha$, i.e. $\alpha=\frac{5}{3}$ in Eq.(\ref{x01}) and $\alpha=\frac{7}{3}$ in Eq.(\ref{x02}), the polarity of $\chi_1$ ($\chi_2$ if $\chi_1=0$) completely determines the positivity, or otherwise, of the tangent of the future directed null geodesics ceasing at the singularity, at $(R,u)=(0,0)$.  In fact it can be seen that the initial conditions causing the first non-zero $\chi_i>0$ will cause the time curve of the apparent horizon to be increasing, leading to the formation of singularity prior to the formation of trapped surfaces, causing a chance for a family of outgoing null geodesics of non-zero measure to leave the central singularity. The time curve of the singularity is given by
\begin{equation}\label{singcurve3}
    t(r,0)=t_s(r)=t_i+\int_{0}^{1} \frac{e^{-\nu}}{\sqrt{\frac{\mathcal{M}}{v}+\frac{be^{2A}-1}{r^2}}} dv,
\end{equation}
from which $\chi _1 (0)$ is calculated as
\begin{equation}\label{chi1}
    \chi _1(0)=-\frac{1}{2}  \int _{0}^{1}\frac{b_{01}}{\bigg (\frac{M_0}{v}+b_{00}+2A_2 \bigg )^{\frac{3}{2}}} dv.
\end{equation}
Here, $b_{00}$, $b_{01}$, etc. are components of Taylor expansion of $b_0$ around the regular center. If $b_{01}\neq 0 $, then the polarity of $b_{01}$ determines the end state. However, if $b_{01}=0$, we move on to investigating $\chi _2$ and based on its polarity, the end state of the collapse can be determined using the expression given by
\begin{widetext}
\begin{equation}\label{chi2}
 \chi _2(0)= \int _{0}^{1}\Bigg [\frac{3}{8}\frac{b_{01}^{2}}{\bigg (\frac{M_0}{v}+b_{00}+2A_2 \bigg )^{\frac{5}{2}}}-\frac{g_{2}}{\sqrt{\frac{M_0}{v}+b_{00}+2A_2}} -\frac{1}{2}\frac{\frac{M_2}{v}+2A_{2}^{2}+2b_{00}A_2+b_{02}}{\bigg (\frac{M_0}{v}+b_{00}+2A_2 \bigg )^{\frac{3}{2}}} \Bigg ] dv.
\end{equation}
\end{widetext}
where 
\begin{equation}\label{g2}
    g_{2}=\frac{1}{2} v A_{2,v}.
\end{equation}

In the next section, we now consider a pressureless (dust) collapse and add a small perturbation in $\mathcal{M}$, which correspondingly gives a small pressure perturbation within the collapsing matter cloud. Our goal is to examine, how such a 
pressure perturbation affects the final state of collapse in terms of either a black hole or naked singularity formation.

\section{Pressure Perturbations in Dust with Inhomogeneity}
In the absence of any EoS, we have five field equations in six unknowns as mentioned in Section II. Hence, we are left with one degree of freedom and thereby allowed to specify how the mass profile $\mathcal{M}$ evolves with the collapse.  In case of zero pressure, i.e. inhomogeneous dust, the corresponding mass profile near the regular center, up to two orders in $r$ is given by
\begin{equation}\label{mpd}
    \mathcal{M}(r,v)=m_0+m_2r^2,
\end{equation}
where $m_0$ and $m_2$ are constants, since if they depend on $v$, then the pressure becomes non-zero necessarily, as is apparent from Eq.(\ref{efe2}).
\begin{figure*}
\subfigure[$\mathcal{M}(r,v)=m_0+m_2 r^2+\gamma(1-v)^{\alpha}r^2$]
{\includegraphics[scale=0.45]{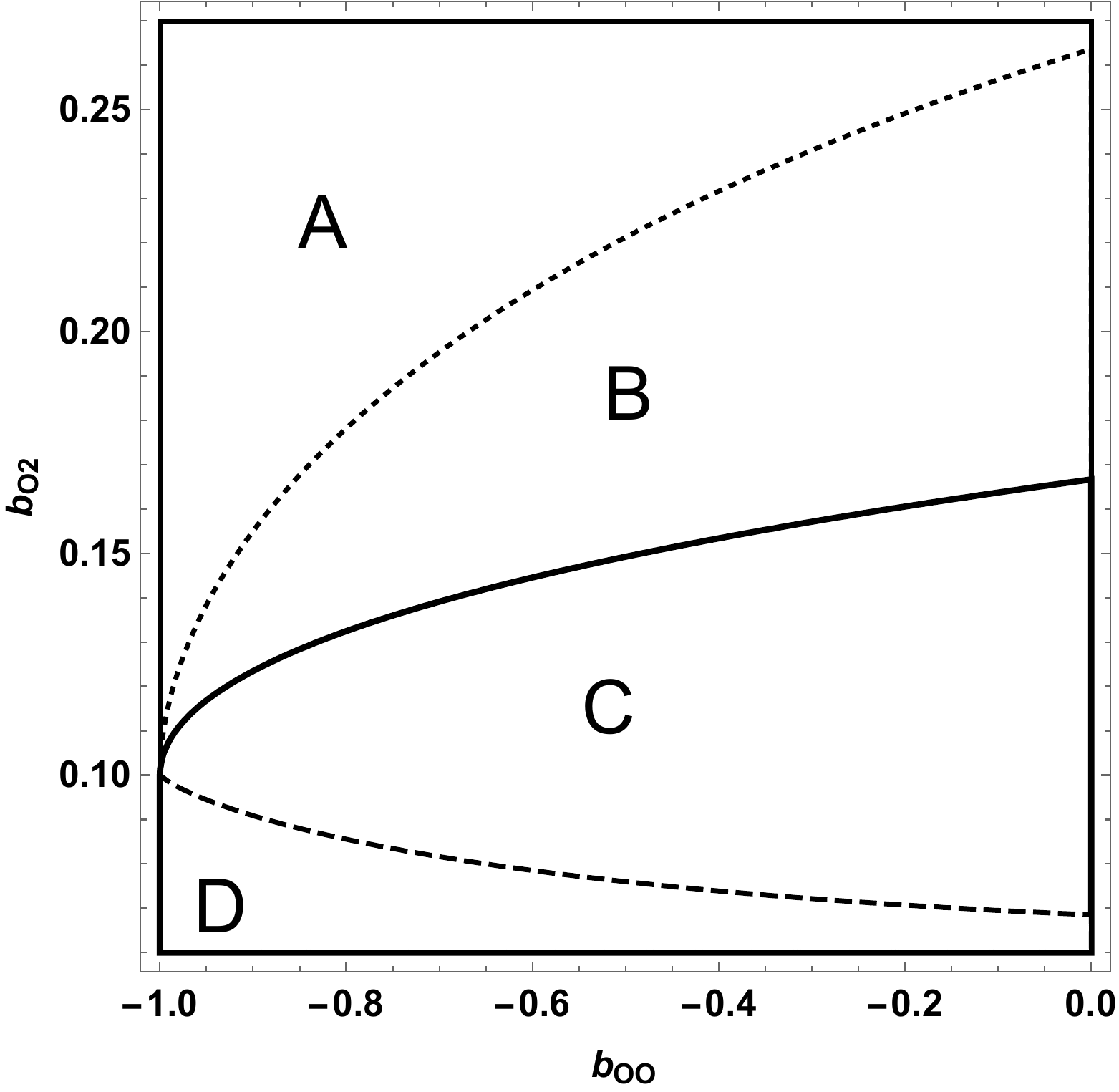}}
\label{1a}
\hspace{0.2cm}
\subfigure[$\mathcal{M}(r,v)=m_0+m_2 r^2+\gamma(v-v^2)r^2$]
{\includegraphics[scale=0.45]{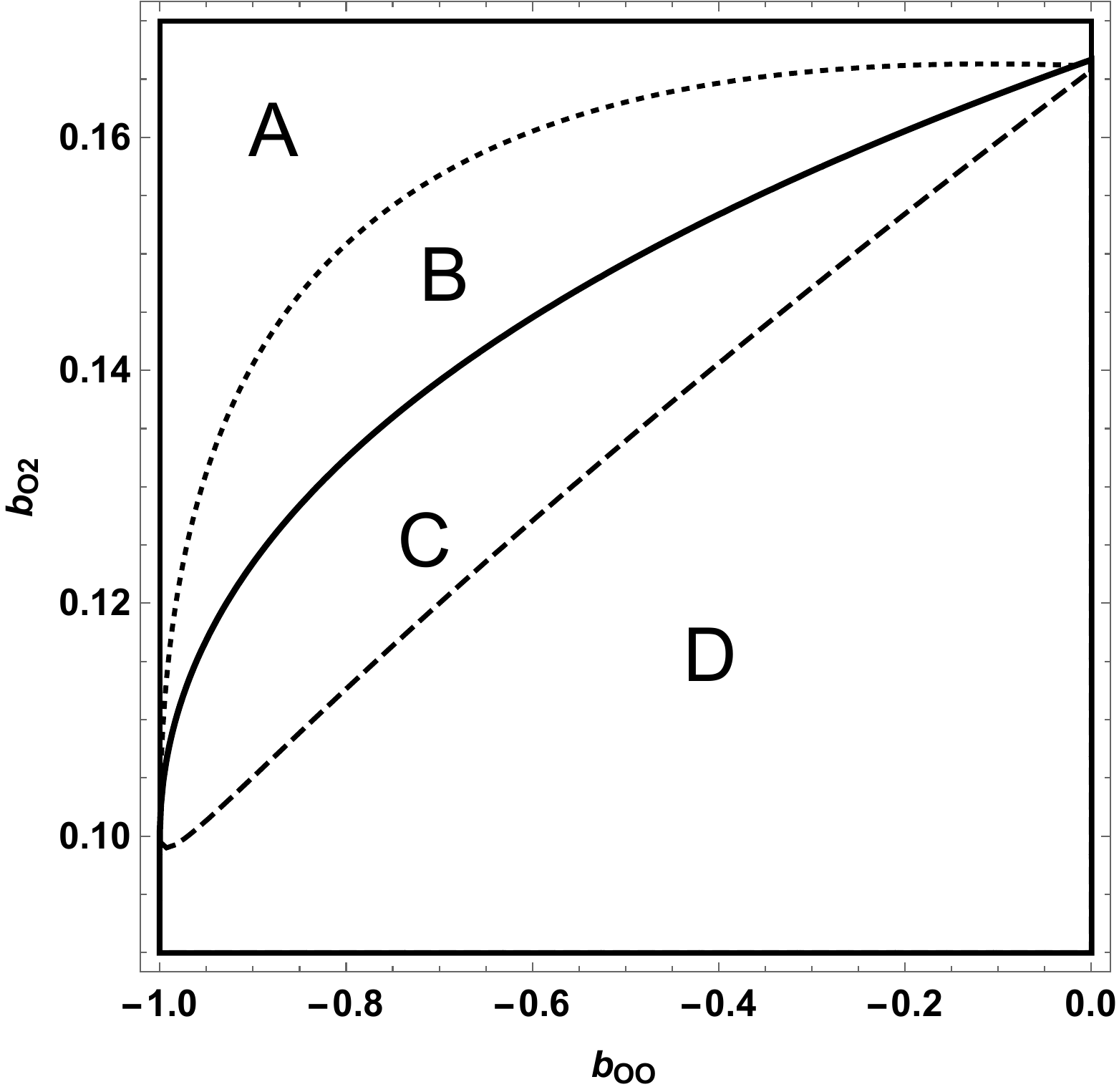}}
\label{1b}
\hspace{0.2cm}
\subfigure[$\mathcal{M}(r,v)=m_0+m_2\beta (v+v^2-v^3) r^2$]
{\includegraphics[scale=0.45]{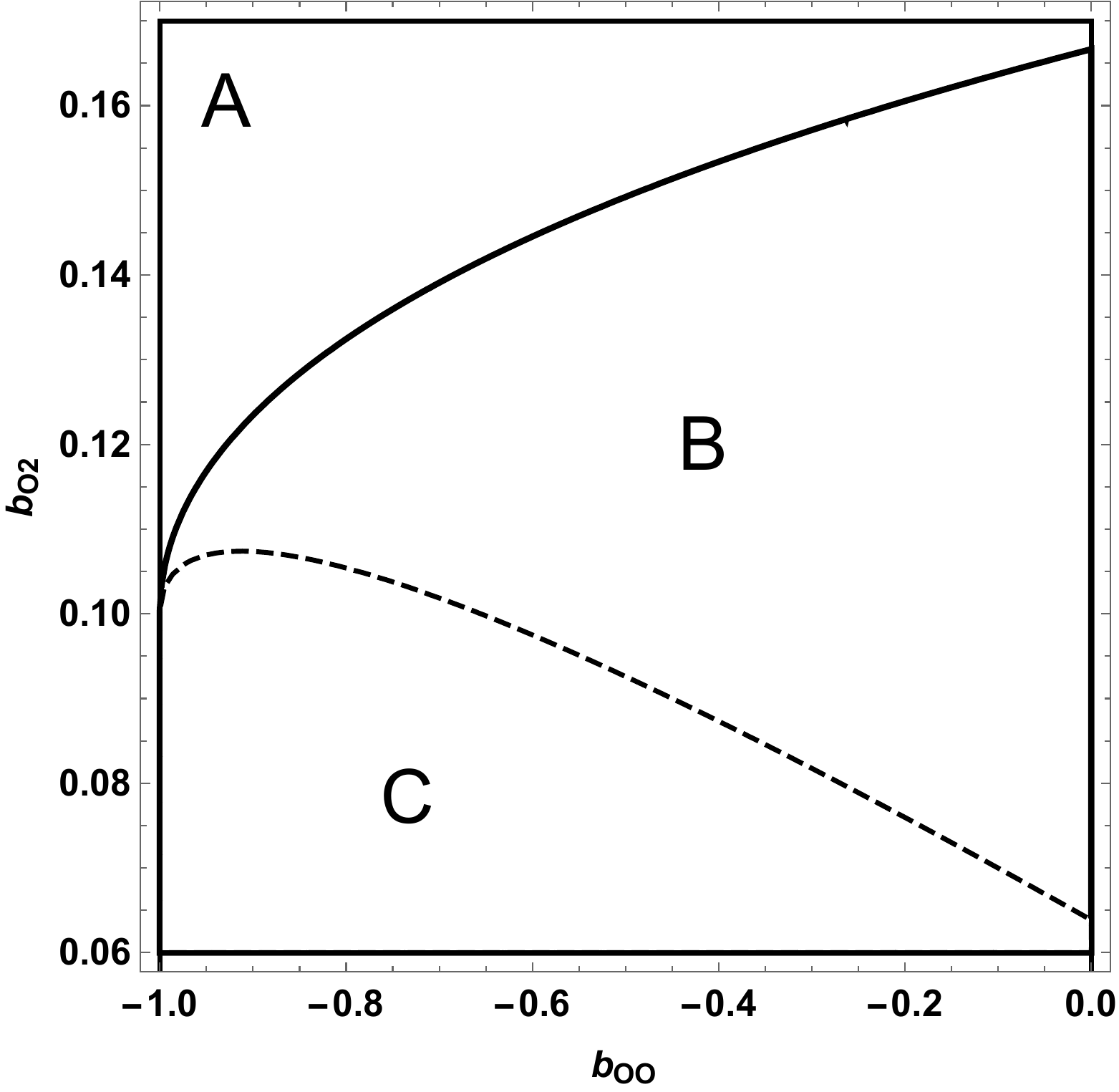}}
\label{1c}
\hspace{0.2cm}
\subfigure[$\mathcal{M}(r,v)=m_0+m_2\beta (v-v^2+v^3) r^2$]
{\includegraphics[scale=0.45]{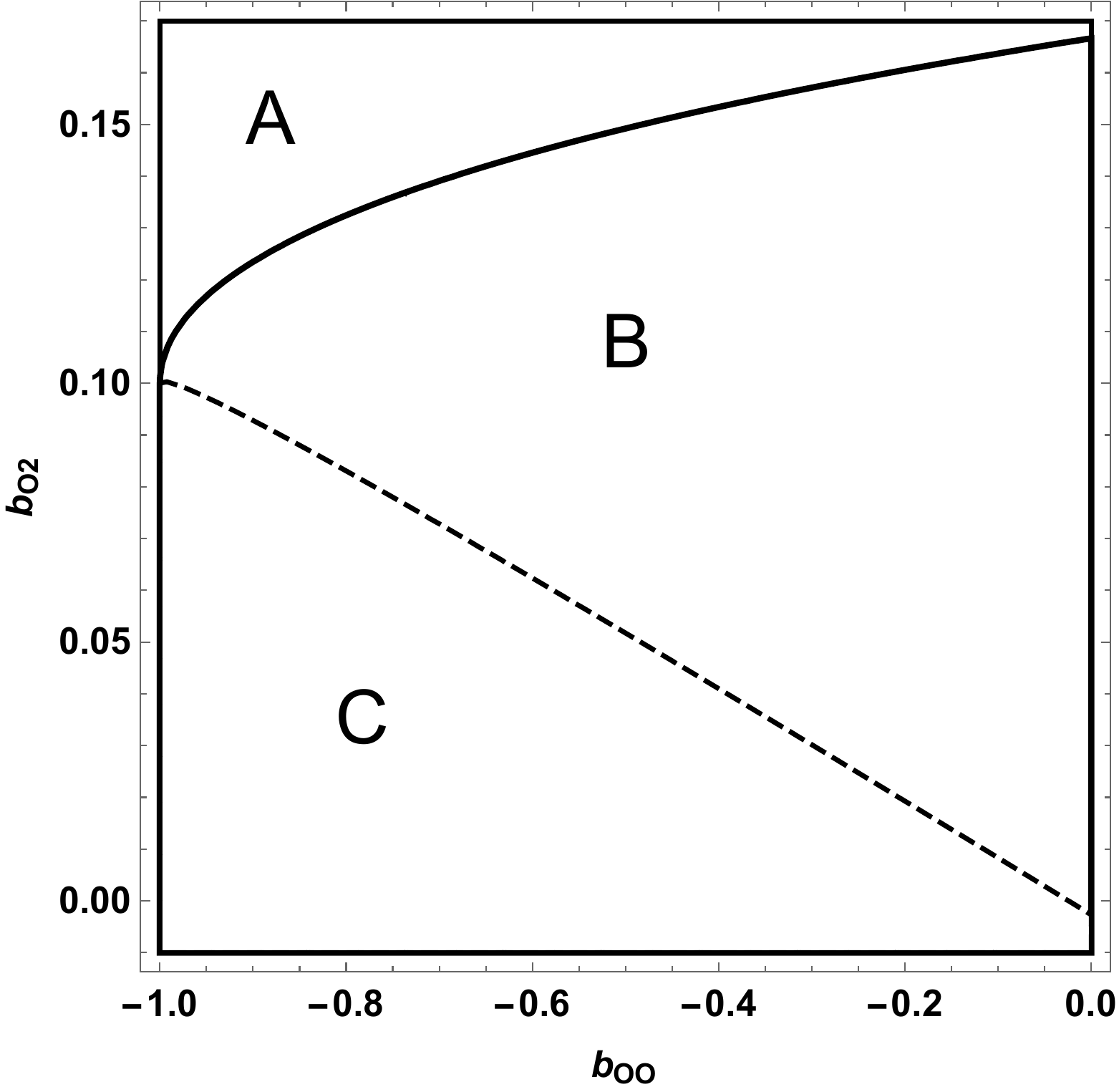}}
\label{1d}
\hspace{0.2cm}
\subfigure[$\mathcal{M}(r,v)=m_0+C_1v^3+C_2+m_2r^2$]
{\includegraphics[scale=0.45]{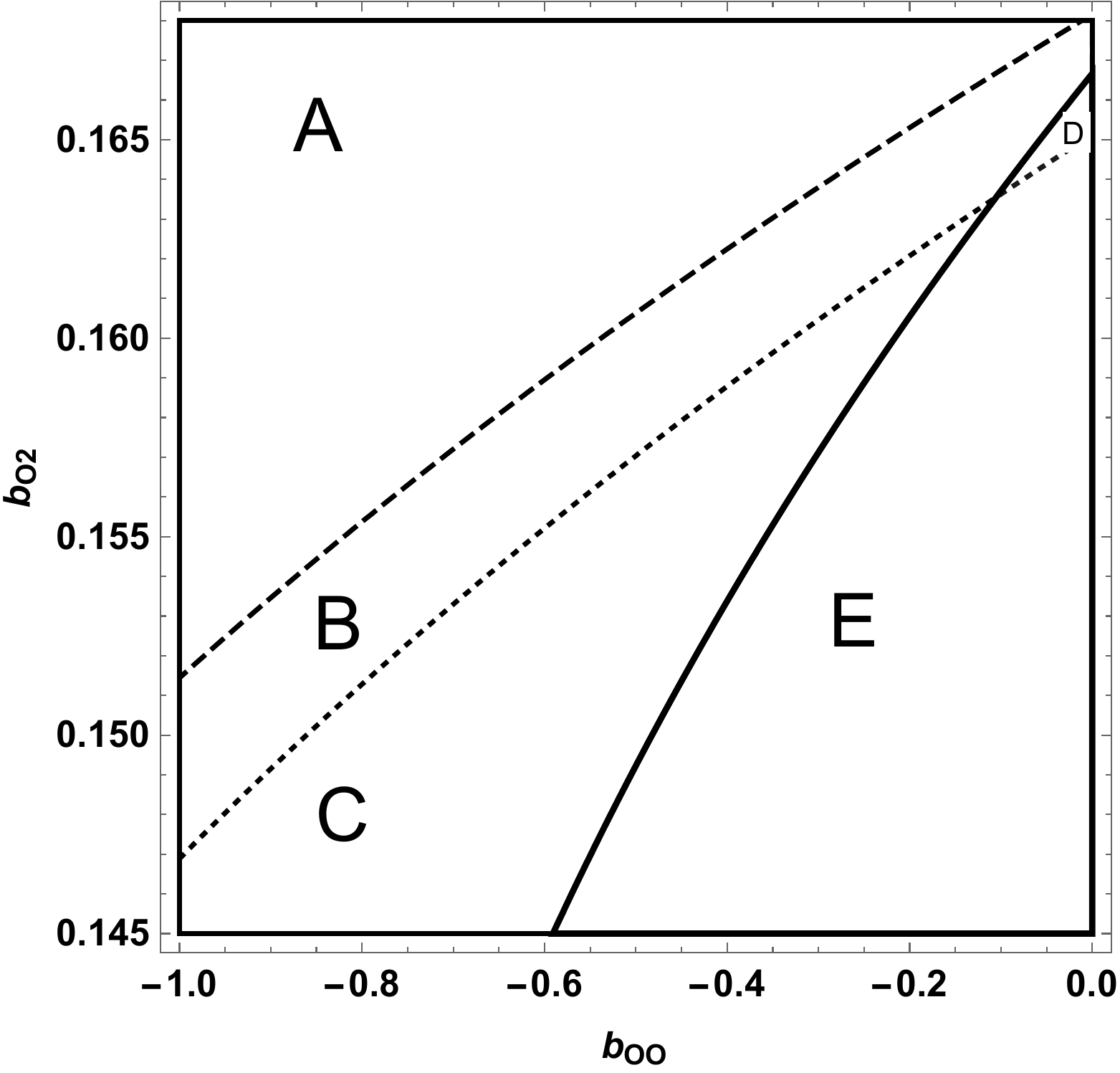}}
\label{1e}
\caption{The horizontal axis is $b_{00}$ and the vertical axis is $b_{02}$. Here $m_0=1$ and $m_2=-0.1$. The entire region above (below) the solid curve gives a black hole (naked singularity) as end state. Adding positive (negative) perturbation to the unperturbed $\mathcal{M}=m_0+m_2r^2$ shifts this curve to the one represented by dashed (dotted) curve indicating that small pressure could change the outcome of the collapse.  $\gamma=0.1(-0.1)$, $\beta=1 (-1)$ and $C_1=0.1(-0.1)$ for positive (negative) perturbation. Also $\alpha=1.5$ in 1.(a) and $C_2=1$ in 1.(e). The regions bounded/separated by the solid, dashed and/or dotted curves are denoted by A, B, C, D and E.}
\label{fig1}
\end{figure*}
\begin{table*}[t!]
\centering
\begin{tabular}{|l|c|c|c|c|c|}
\hline
 $\delta_1=\gamma(1-v)^{\alpha}r^2$  & Dust &  $\gamma=0.1$ &  $\gamma=-0.1$\\
\hline
\hline
Black Hole:                           & A, B       & A, B, C       &  A             \\
\hline 
Naked Singularity:                   & C, D       & D             & B, C, D       \\
\hline
\end{tabular}
\quad
\begin{tabular}{|l|c|c|c|c|c|}
\hline
 $\delta_2=\beta (v+v^2-v^3) r^2$     & Dust  &  $\beta=1$   \\
\hline
\hline
Black Hole:                           & A,        &  A, B               \\
\hline 
Naked Singularity:                    & B, C     & C               \\
\hline
\end{tabular}

\vspace{\baselineskip}

\begin{tabular}{|l|c|c|c|c|c|}
\hline
$\delta_1=\gamma(v-v^2)r^2$        & Dust &  $\gamma=0.1$ & $\gamma=-0.1$ \\
\hline
\hline
Black Hole:                        & A, B       & A, B, C       &  A             \\
\hline 
Naked Singularity:                  & C, D       & D             & B, C, D      \\ 
\hline
\end{tabular}
\quad
\begin{tabular}{|l|c|c|c|c|c|}
\hline
 $\delta_2=\beta (v-v^2+v^3) r^2$ & Dust &  $\beta=1$ \\
\hline
\hline
Black Hole:                          & A           & A, B                      \\
\hline 
Naked Singularity:                   & B, C       & C                         \\
\hline
\end{tabular}

\vspace{\baselineskip}

\begin{tabular}{|l|c|c|c|c|c|}
\hline
 $\delta_3=C_1v^3+C_2$      &   Dust       & $C_1=0.1$, $C_2=1$      & $C_1=-0.1$, $C_2=1$   \\
\hline
\hline                        
Black Hole:                    & A, B, C                          & A                 & A, B. D         \\
\hline 
 Naked Singularity:            &  D, E                        & B, C, D, E        & C, E          \\
\hline
\end{tabular}
\caption{Classification of different regions in the $(b_{00},b_{02})$ plane on the basis of the outcome of its end state, for different perturbations corresponding to each of the four perturbations. Here $\alpha=1.5$ in the first perturbation.}
\label{tab1}
\end{table*}

\subsection{Different types of perturbations }
We consider here three types of perturbations to the mass profile corresponding to inhomogeneous dust, Eq.(\ref{mpd}). This gives the modified, or perturbed mass profile as follows:
\begin{enumerate}
  
    \item $\mathcal{M}(r,v)=m_0+m_2r^2+\delta_1 (v)r^2$.

The perturbed part, $\delta_1(v)$, is minimally coupled to the component $m_2$ of order ($r^2$) of the Taylor expansion of mass profile of dust. The expressions for the functions required to determine the outcome of end state of collapse corresponding to this $\mathcal{M}$, hereinafter referred to as type 1 function, could be obtained from Eq.(\ref{A2int}) as follows:
\begin{equation} \label{A21g21}
A_{2_{I}}=-\frac{2}{3 m_0}\delta_1, \hspace{0.25cm}   g_{2_{I}}=-\frac{v}{3m_0}\delta_{1,v}. 
\end{equation}
For $(r,v)$ close to $(0,0)$, $R'= rv'+v$ can be approximated as $R' \sim 2v$, where we have used L'Hospital's rule to approximate $v'$ as $\frac{v}{r}$. For $r$ close to zero and $v>>0$, we have $R'\sim v$.   The density and  pressure profiles for this mass function approximated for very small $r$ but large $v$, are obtained from Eq.(\ref{efe1}) and Eq.(\ref{efe2}) respectively, and  are  given by
\begin{equation} \label{d1p1}
\rho_{I} = \frac{3m_0}{v^3}+5\frac{m_2+\delta_1}{v^3}r^2,  \hspace{0.25cm}  p_{I}=-\frac{\delta_{1,v}}{v^2}r^2. 
\end{equation}
The formulation for density is valid only for very early phases of collapse, and not near the time of formation of the singularity, since $R'$ is approximated by $v$. The linear EoS parameter, which is the ratio of pressure and density, turns out to be 
\begin{equation} \label{k1}
k_I(r,v)=-\frac{\delta_{1,v}vr^2}{3m_0+5(m_2+\delta_1)r^2}.
\end{equation}

\item $\mathcal{M}(r,v)=m_0+m_2\delta_2 (v)r^2$.

The perturbed part, $\delta_2(v)$ is non-minimally coupled to the component $m_2$ of the unperturbed mass profile.  On similar lines as the previous case, the  expressions for the functions required to determine the outcome of the end state of collapse corresponding to this $\mathcal{M}$, hereinafter referred to as type 2 function, is obtained from Eq.(\ref{A2int}) as follows:
 \begin{equation}\label{A22g22}
     A_{2_{II}}=-\frac{2m_2}{3 m_0}(1-\delta_2), \hspace{0.25cm}  g_{2_{II}}=-\frac{v m_2}{3m_0} \delta_{2,v}.
\end{equation}
 The density and pressure profile for this mass profile are obtained using Eq.(\ref{efe1}) and Eq.(\ref{efe2}) respectively, and given by
\begin{equation}\label{d2p2}
\rho_{II} = \frac{3m_0}{v^3}+5\frac{m_2\delta_2}{v^3}r^2, \hspace{0.25cm} p_{II}=-\frac{m_2\delta_{2,v}}{v^2}r^2.
\end{equation}        
Here as well, $R'$ is approximated by $v$  so that the approximation for density holds good for $v>>0$ and $r$ near zero. The linear EoS parameter is 
\begin{equation}
    k_{II}=-\frac{m_2\delta_{2,v}vr^2}{3m_0+5m_2\delta_{2}r^2}.
\end{equation}
\begin{figure*}
\centering
\subfigure[$\mathcal{M}(r,v)=m_0+m_2 r^2+\gamma(1-v)^{\alpha}r^2$]
{\includegraphics[scale=0.5]{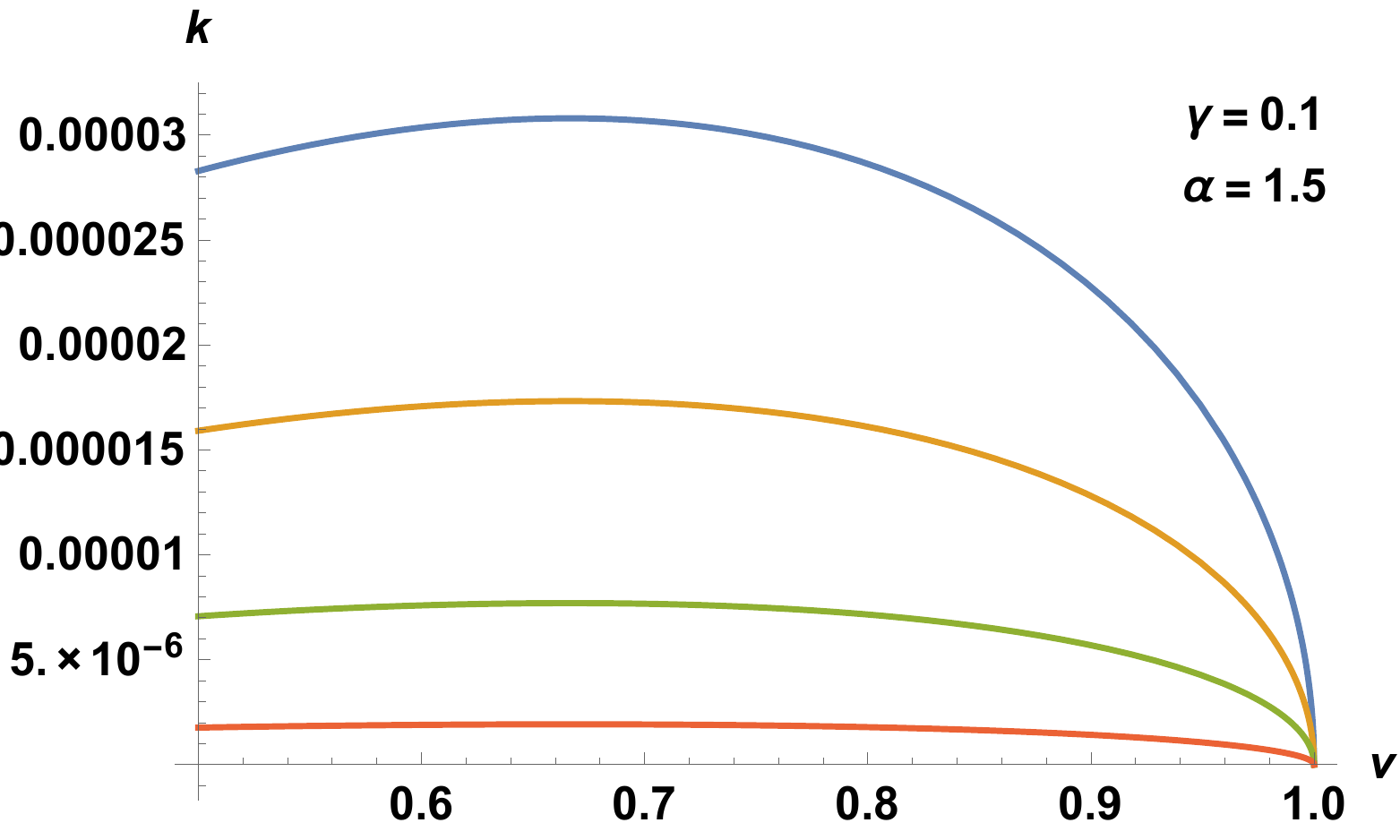}}
\label{2a}
\hspace{0.2cm}
\subfigure[$\mathcal{M}(r,v)=m_0+m_2 r^2+\gamma(1-v)^{\alpha}r^2$]
{\includegraphics[scale=0.5]{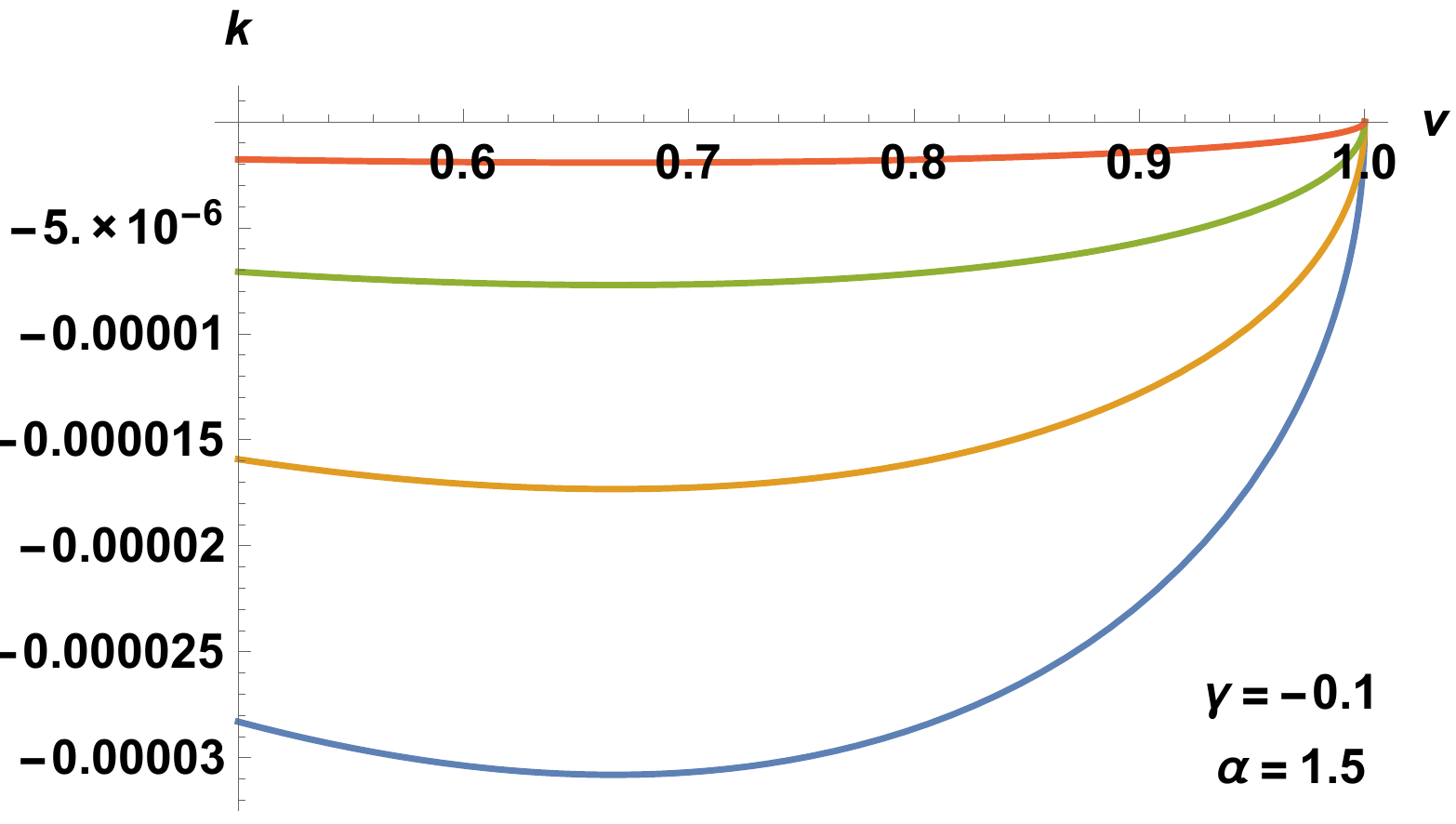}}
\label{2b}
\hspace{0.2cm}   
\subfigure[$\mathcal{M}(r,v)=m_0+m_2 r^2+\gamma(v-v^2)r^2$]
{\includegraphics[scale=0.5]{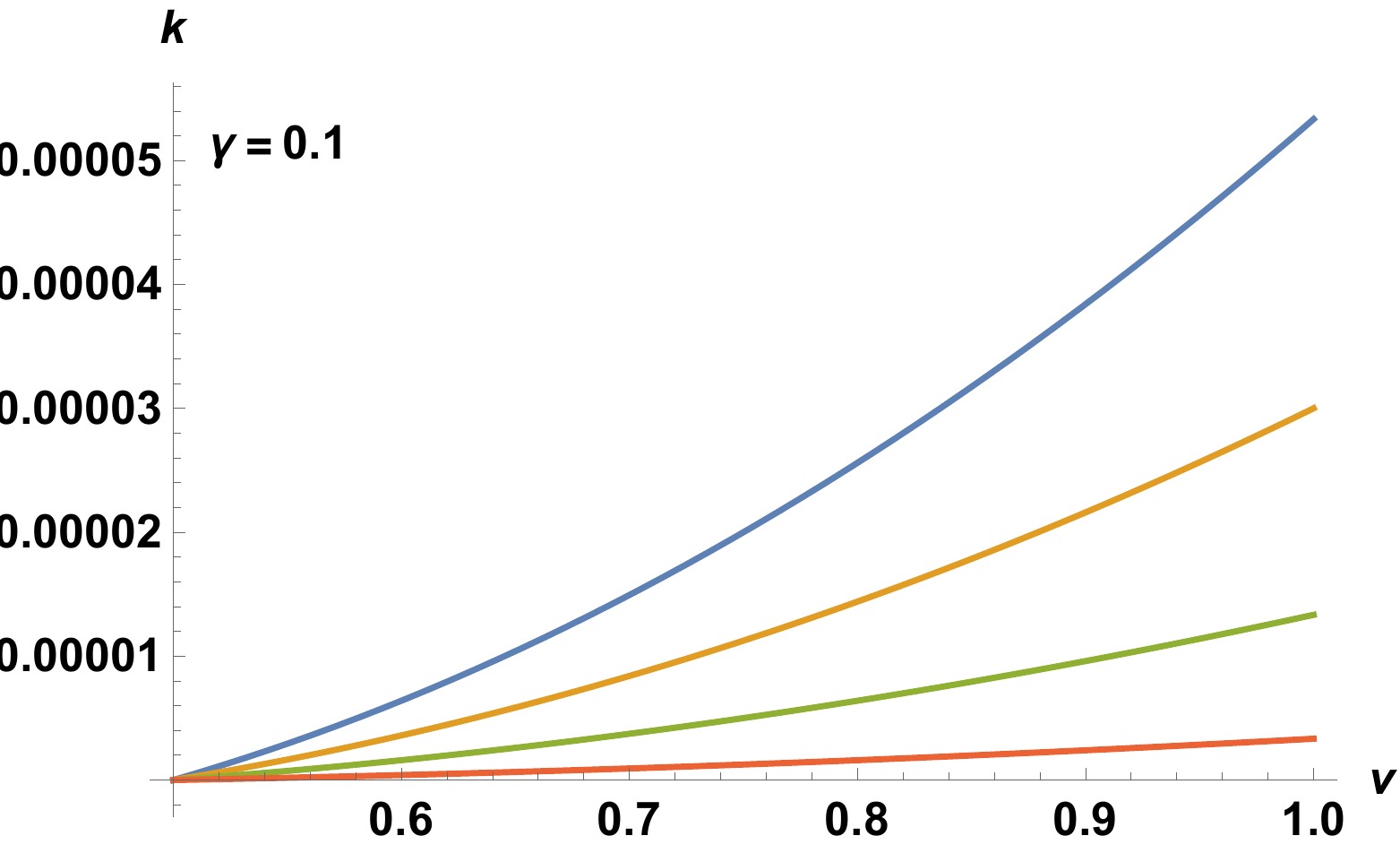}}
\label{2c}
\hspace{0.2cm}
\subfigure[$\mathcal{M}(r,v)=m_0+m_2 r^2+\gamma(v-v^2)r^2$]
{\includegraphics[scale=0.5]{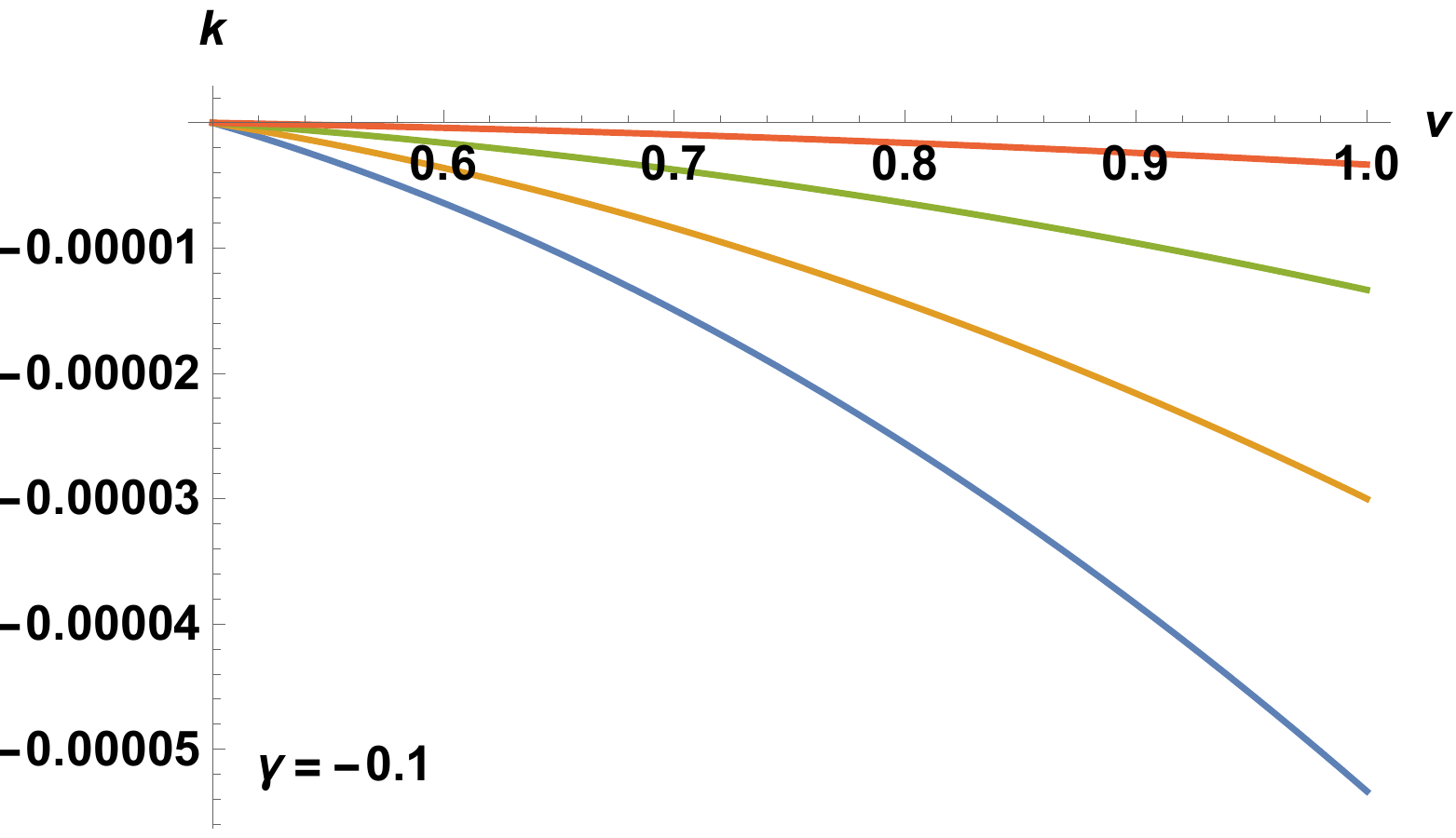}}
\label{2d}
\hspace{0.2cm}
\subfigure[$\mathcal{M}(r,v)=m_0+m_2\beta (v+v^2-v^3) r^2$]
{\includegraphics*[scale=0.5]{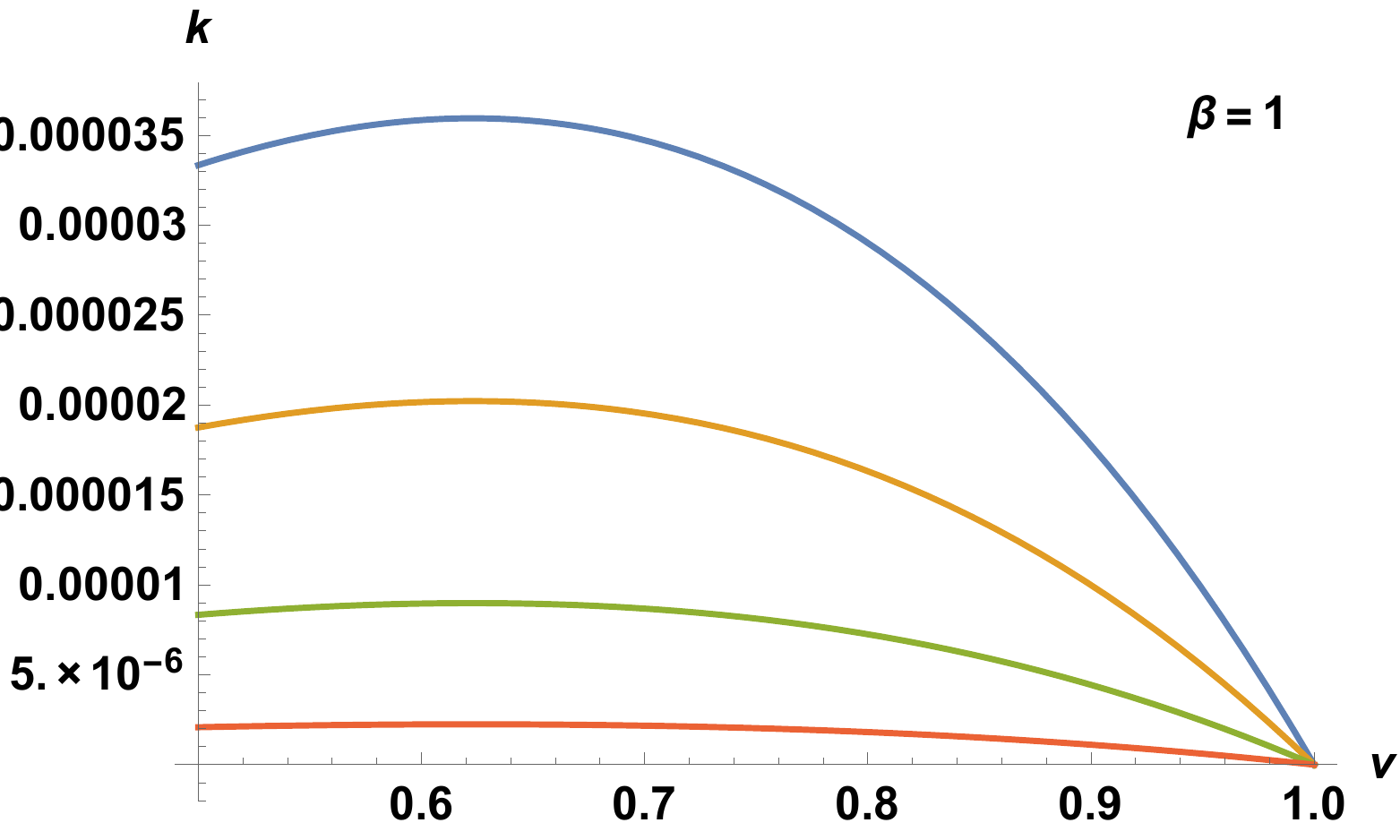}}
\label{2e}
\hspace{0.2cm}
\subfigure[$\mathcal{M}(r,v)=m_0+m_2\beta (v-v^2+v^3) r^2$]
{\includegraphics[scale=0.5]{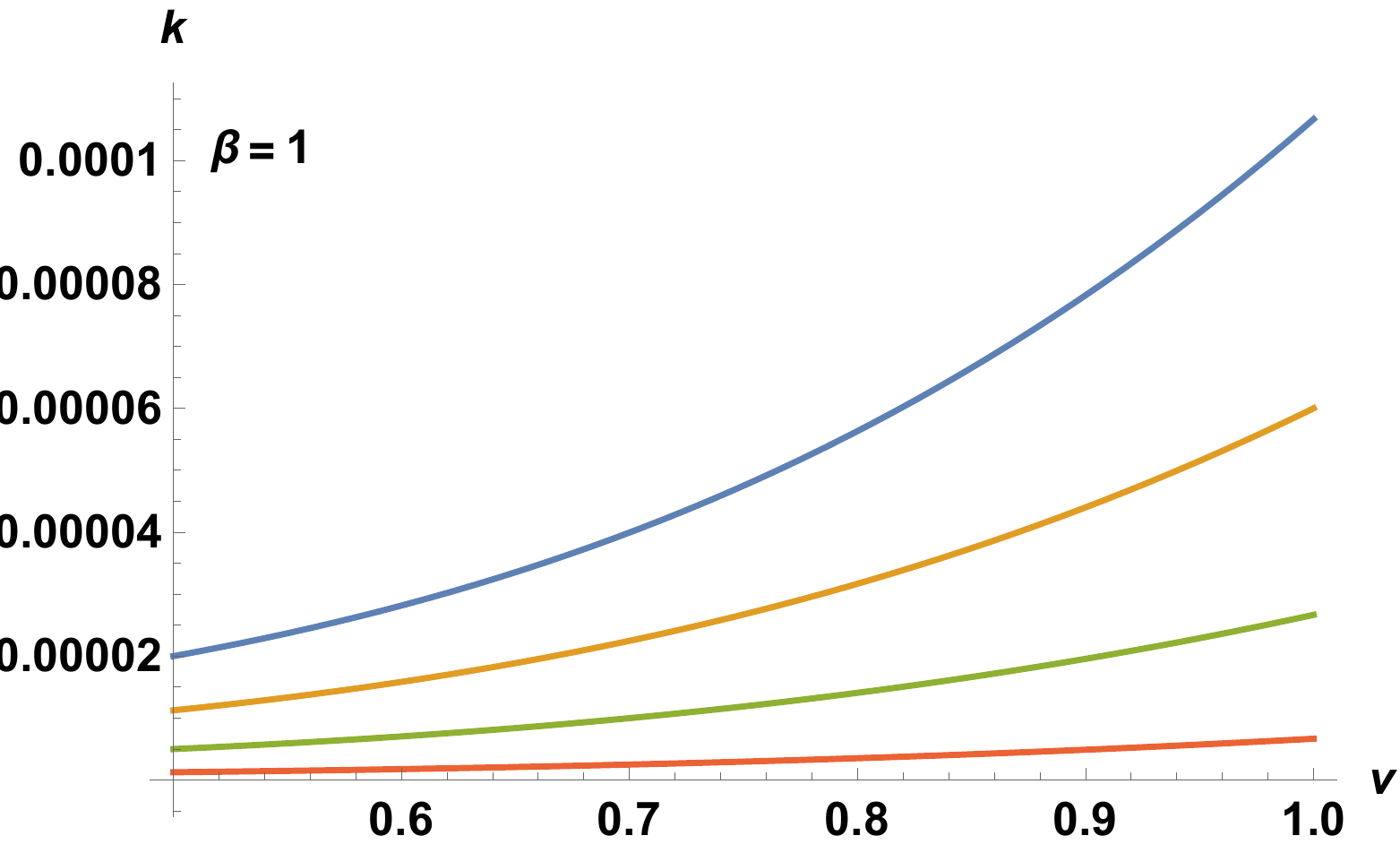}}
\label{2f}
\hspace{0.2cm}
\subfigure[$\mathcal{M}(r,v)=m_0+C_1v^3+C_2+m_2r^2$]
{\includegraphics[scale=0.5]{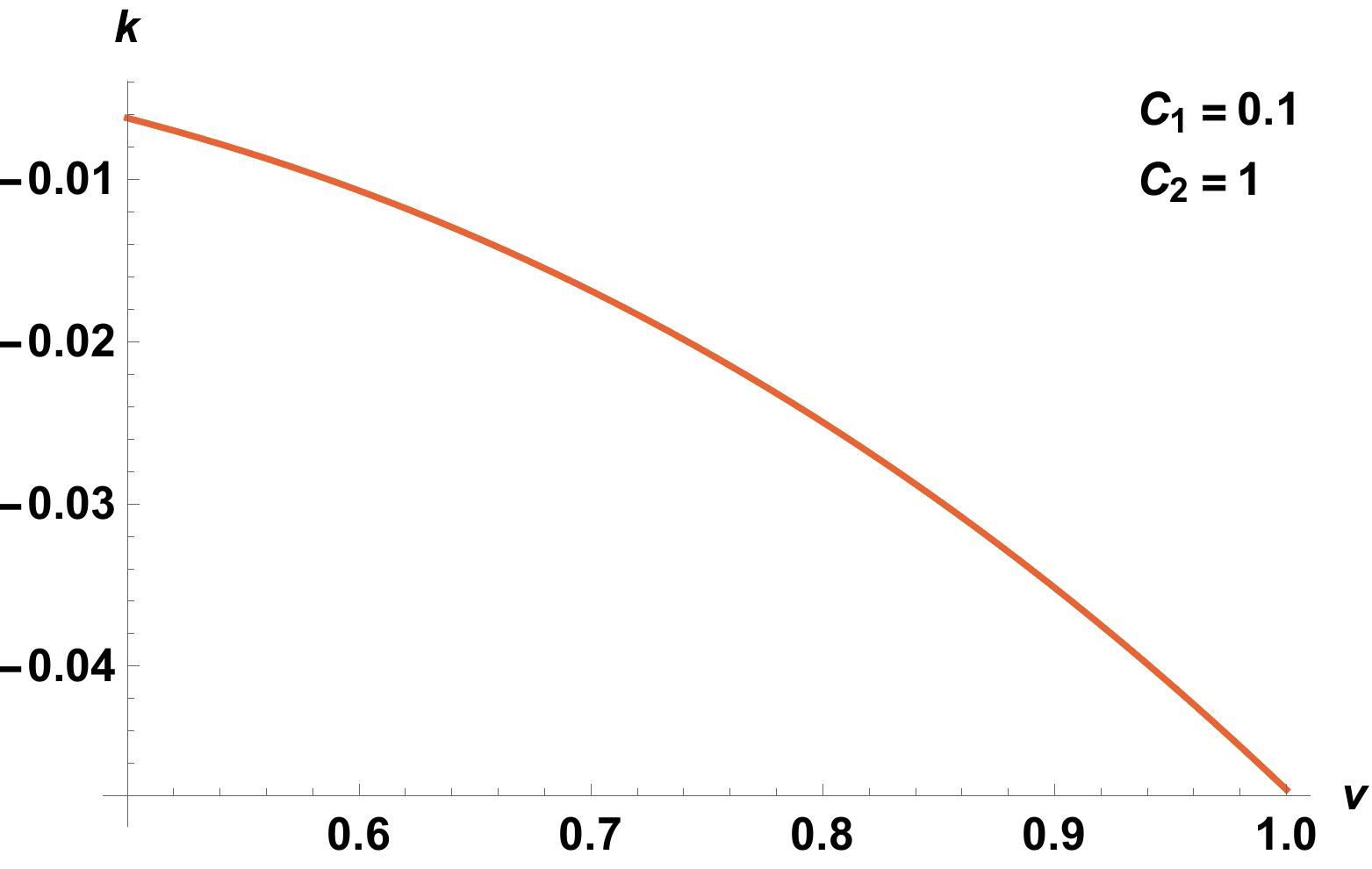}}
\label{2g}
\hspace{0.2cm}
\subfigure[$\mathcal{M}(r,v)=m_0+C_1v^3+C_2+m_2r^2$]
{\includegraphics[scale=0.5]{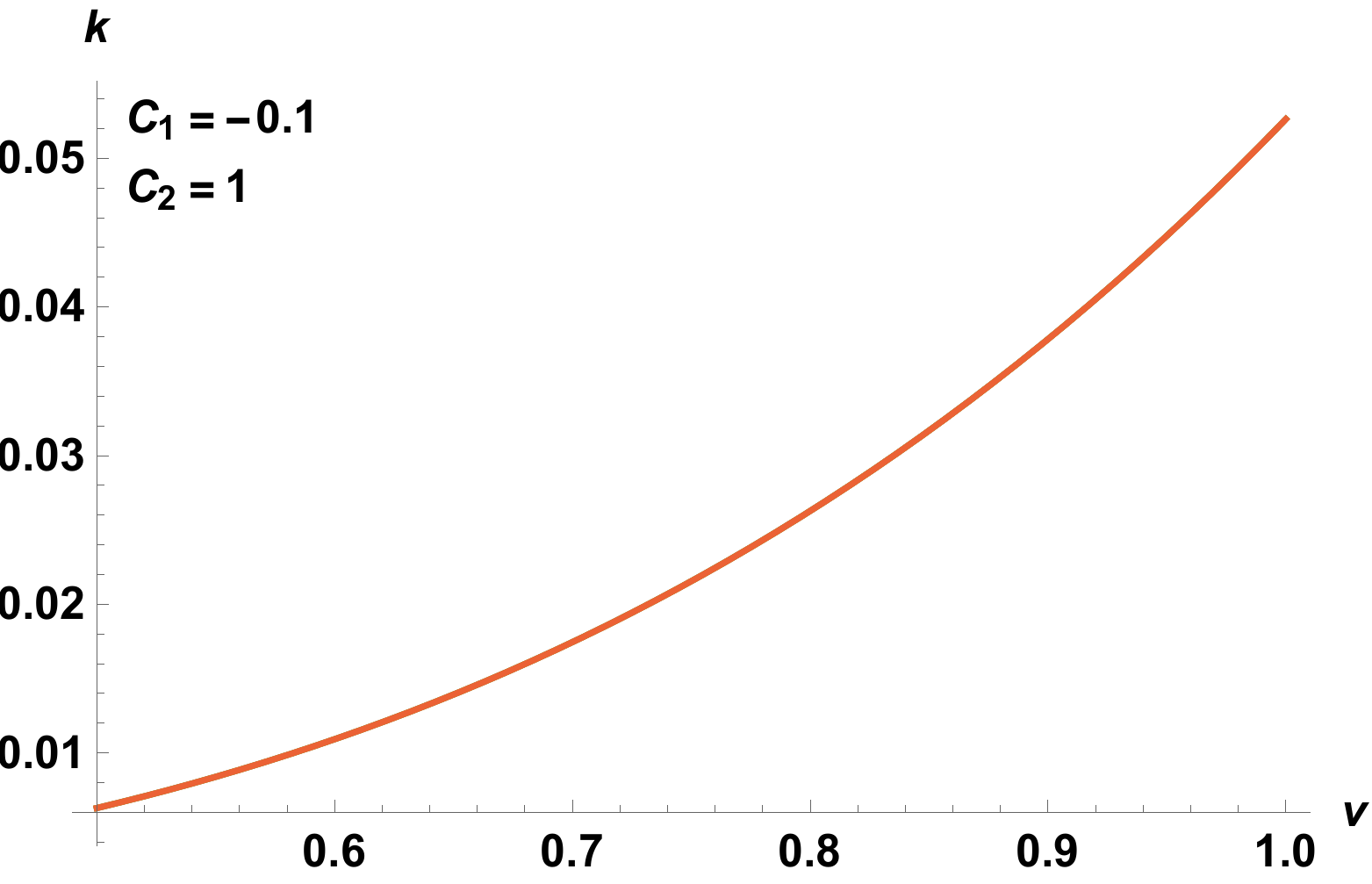}}
\label{2h}
\hspace{0.2cm}
 \caption{Dynamics of EoS parameter ($k$) with the transformed co-ordinate ($v$) for mass profiles corresponding to different types of perturbations,  as mentioned in the sub-figures, and corresponding to different radial co-ordinates ($r$) taking values $0.01$, $0.02$, $0.03$ and $0.04$, represented by red, green, yellow and blue colors respectively. Since the approximation $R'\sim v$ has been used, the formula for EoS parameter holds true only for $v$ far from zero and $r$ close to zero.}
\label{fig2}
\end{figure*}

\item $\mathcal{M}=m_0+\delta _3(v)+m_2 r^2$.

The perturbation $\delta_3(v)$,  given to the component $m_0$ of mass profile corresponding to inhomogeneous dust  Eq.(\ref{mpd}), causes an obstruction in calculating $A_2$ because of the presence of $v'_{,r}$ term in the numerator of Eq.(\ref{A2int}), which demands the knowledge of the function $v(t,r)$, or in other words, the dynamics of collapse, which is not known in a general scenario.  A particular case for which the requirement of $v$ could be avoided is when $M_{0,vv}=\frac{2M_{0,v}}{v}$. Such a condition can be achieved by choosing $\delta_3(v)$ as
\begin{equation}
    \delta _3 (v)= C_1 v^3+C_2,
\end{equation}
giving us the mass profile hereinafter referred to as type 3 perturbation. $A_2$ corresponding to such perturbed mass profile vanishes, and so does $g_2$.  The density for $v>>0$ and $r$ close to zero, and pressure profile is obtained using Eq.(\ref{efe1}) and Eq.(\ref{efe2}) respectively, and is given by
\begin{equation}\label{d2p2}
\rho_{III} = \frac{3(m_0+C_1v^3+C_2)+5m_2r^2}{v^3},  \hspace{0.25cm} p_{III}=-3C_1.
\end{equation}        
The linear EoS parameter is 
\begin{equation}
    k_{III}=\frac{3C_1v^3}{3(m_0+C_1v^3+C_2)+5m_2r^2}.
\end{equation}

\end{enumerate}

In all the three types, once we get the value for $A_2$ for each case, we use it to find the polarity of $\chi_2$, provided $b_{01}$ is equal to zero, by using Eq.(\ref{chi2}). Along with $A_2$, we also substitute in this equation, the values $M_0$ and $M_2$, which in these cases are components of Taylor expansion of perturbed mass profile (for example, in type 1 perturbation, we have $M_0=m_0$ and $M_2=m_2+\delta_1(v)$). For $b_{01}\neq 0$, the polarity of $\chi_1$ is required to determine the end-state of collapse, which is based on the polarity of $b_{01}$ only. This is because, $\frac{M_0}{v}+b_{00}+2A_2>0$ always, to avoid the emergence of imaginary term  in Eq.(\ref{chi1}). This means that adding pressure perturbation to the pressureless collapse does not cause any change in the outcome of the end state if $b_{01}\neq 0$.  

One could alternatively think about starting with a homogeneous perfect fluid (non-zero pressure) and then bringing it closer to a more realistic model by adding a perturbation such that it makes the perfect fluid inhomogeneous.  In this case, the unperturbed mass profile is a function of time only, i.e. $\mathcal{M}=\mathcal{M}(t)$, nevertheless, in the $(r,v)$ coordinates, it is a function of both $r$ and $v$, making the components $m_0^{*}$ and $m_2^{*}$ of the Taylor expansion of $\mathcal{M}$ around $r$ to be functions of $v$ instead of being constants as in the case of zero pressure. This makes the integral Eq.(\ref{A2int}) difficult to solve due to the presence of $v'_{,r}$ term, unless either $M_0$ is constant or at least $M_{0,vv}=\frac{2M_{0,v}}{v}$ similar to what we saw in case of type 3 perturbation. In case of homogeneous dust collapse, $v=v(t)$ and hence $v'=0$, which if used in Eq.(\ref{A2int}) causes the second term in the numerator to vanish. However, the question is: Can we substitute this in Eq.(\ref{A2int}) along with substituting the components $M_0$ and $M_2$ of the perturbed mass profile? 
The answer to this is not in affirmative. This is because the perturbed fluid is no longer homogeneous and hence $v=v(t)$ does not hold anymore. 

It should be noted that in the case of giving perturbation in the mass profile of an inhomogeneous dust, causing perturbation of a non-zero pressure, i.e. giving type 1, type 2 or type 3 perturbations, no such problem is faced.

\subsection{Observations}
\begin{itemize}
\item The results obtained are compressed and depicted in Fig.(\ref{fig1}) for five different expressions of perturbation corresponding to type 1, type 2 and type 3 in $\mathcal{M}$. It depicts the changes obtained in the outcome of the end state due to the addition of these perturbations (one at a time). This can be investigated by checking if for a given value of initial condition, in this case $(b_{00},b_{02})$, i.e. the components of the Taylor expansion of velocity profile around regular center, whether or not the polarity of $\chi_2$ is affected due to consideration of perturbed mass profile as compared to the unperturbed (inhomogeneous dust) mass profile. As an example, in Fig.(1a), in absence of any perturbation, the region $(b_{00},b_{02})$ is divided by a solid curve on the basis of polarity of $\chi_2$. Values of $(b_{00},b_{02})$ above this curve when substituted in Eq.(\ref{chi2}) give negative value of $\chi_2$, while those points below the solid curve give positive values of $\chi_2$. Hence regions $A$ and $B$ give the endstate as a black hole while the regions $C$ and $D$ give the end state as a naked singularity. Adding perturbation to this mass profile  will change the way the region $(b_{00},b_{02})$ is divided. In presence of negative type 1 perturbation, the dotted curve in Fig.(1a) separates this region in such a way that all points above this curve, i.e. region $A$ ends up in a black hole while all points below it, i.e. regions $B$, $C$, and $D$ end up in a naked singularity. Hence we could see that region $B$, which ended up in a black hole now ends up to be a naked singularity in the presence of negative type 1 perturbation. Similarly, other figures could be understood.    
 
\item Table ($1$) presents the classification of regions in the $(b_{00},b_{02})$ plane based on whether the end state turns out to be a black hole or a naked singularity, for each form of perturbations mentioned.

\item We demand the matter field obtained due to the perturbation $\delta(v)$ in $\mathcal{M}$ to satisfy at least the weak energy condition, which in the case of a perfect fluid is expressed as
\begin{equation}
    \rho \geq 0; \hspace{2cm} \rho+p \geq 0.
\end{equation}
This could be confirmed from Fig.(\ref{fig2}), which is a plot for the dynamics of the EoS parameter $k$  as far as $v>>0$, for different radial coordinates $r$ close to zero. It is assured that $k>-1$ always and in all the perturbations taken into consideration, at least during the initiation of the collapse. The polarity of  pressure could be obtained from the polarity of $k$ as seen in Fig.(\ref{fig2}). It could be equivalently shown that even near the time of formation of singularity, i.e. $v$ near zero, the weak energy condition holds. It is also observed that the satisfaction of weak energy condition restricts the coefficients of the perturbed part in the mass profile, i.e. the coefficients $\gamma$, $\alpha$, $\beta$, $C_1$ and $C_2$. 
\end{itemize}


\section{Concluding Remarks}
Here, we  discuss the concluding points and also note some open concerns:

\begin{enumerate}

    \item Presence of linear EoS $p=k\rho$ with constant $k$ puts a severe restriction on the expression which the mass profile $\mathcal{M}$, satisfying regularity conditions, could take. The restriction is such that the partial differential equation in $\mathcal{M}$, Eq.(\ref{ppde}) must  be compatible with one of its characteristic equations, namely, the partial differential Eq.(\ref{spde}). It should be noted that in the absence of any EoS, the field equations have an additional  degree of freedom and hence freedom of choice for the expression of mass profile (or equivalently the mass function). 
     
    \item Some examples following linear EoS, like inhomogeneous dust, self-similarity, and homogeneous perfect fluid, have their mass profiles such that their corresponding partial differential Eq.(\ref{ppde}) can be reducible to simpler forms. In such cases, the collapse endstates can be worked out clearly.
    
    \item For those mass profiles for which the Eq.(\ref{ppde}) is irreducible to a simpler form, $k$ has to be a variable for $\mathcal{M}$ to satisfy Einstein's field equations.  


    \item Collapsing matter clouds formed due to perturbation in the mass profile corresponding to inhomogeneous dust has a non-zero pressure, making the cloud more realistic compared to inhomogeneous dust which has zero pressure. The matter cloud thus created has a variable EoS parameter $k$. Care has been taken to maintain the weak energy condition of the newly formed matter cloud. Providing the perturbation may or may not change the end state of collapse, depending on its type and  velocity profile of the collapse. The motivation to do the perturbation analysis here is to show that at least as long as no linear EoS is imposed on the matter field, no in-compatibility issue between its corresponding Eq.(\ref{ppde}) and Eq.(\ref{spde}) arises, which is unlike what we get when $k$ is restricted to be a constant. Hence we can proceed to investigate the nature of singularity thus formed.
    
    \item It is important to note that to avoid the usage of the dynamics of collapse, i.e. the information about the scaling function, some restriction has to be imposed on the type of perturbation thus provided. For instance, if the perturbation is coupled to the zeroth component of the Taylor expansion of mass profile of the inhomogeneous dust, the resulting mass profile should have the property that $M_{0,vv}=\frac{2M_{0,v}}{v}$. This limitation is due to the absence of mathematical tools to deal with a more general perturbation. The very reason we take the approach of adding pressure perturbation to dust instead of starting with an arbitrary pressure in investigating the end state of the collapse is the lack of knowledge of the behavior of the scaling function.
   
    \item We also highlight the fact that only the local nakedness of the singularity has been investigated, i.e. the question of whether a future-directed null geodesic can escape the central singularity is answered. Whether or not this geodesic can escape the boundary of the cloud is not discussed and is an open problem which is needed to be investigated separately in a cloud having non-zero pressure. Since the scale of the cloud is kept arbitrary, if the size of the collapsing cloud is considered very large, even a locally visible singularity can be visible to observers close to the singularity. Hence, local nakedness, provided it is generic, should not be taken lightly and can be considered as significant defiance of cosmic censorship.
   
\end{enumerate}


\section{Acknowledgement}
KM would like to thank PSJ for giving him the opportunity to visit the International Center for Cosmology, India. KM also wishes to acknowledge the support of the Council of Scientific and Industrial Research (CSIR, India, Ref: 09/919(0031)/2017-EMR-1), for carrying out this research work.


\begin{thebibliography}{}
 
\bibitem{penrose} R. Penrose, Riv. Nuovo Cimento Soc. Ital. Fis. \textbf{1}, 252 (1969).

\bibitem{joshi}  P. S. Joshi and D. Malafarina, Int. J. Mod. Phys. D \textbf{20}, 2641 (2011).

\bibitem{hawking} S. W. Hawking, G. F. R. Ellis, The large scale structure of space-time, Cambridge University Press (1973).

\bibitem{oppenheimer} J. Oppenheimer and H. Snyder, Phys. Rev. \textbf{56}, 455 (1939).

\bibitem{datt} S. Datt,  \textit{Zs. f. Phys.} \textbf{108}, 314 (1938).  

\bibitem{joshi2} P. S. Joshi and I. H. Dwivedi, Phys. Rev. D \textbf{47}, 5357 (1993). 

\bibitem{mena} F. C. Mena, R. Tavakol and P. S. Joshi, Phys. Rev. D \textbf{62}, 044001 (2000).

\bibitem{dwivedi} I. H. Dwivedi, P. S. Joshi, Class. Quant. Grav. \textbf{9}, L69 (1992). 

\bibitem{deshinkar} S. S. Deshinkar, S. Jhingan, P.S. Joshi, Gen. Relativ. Gravit. \textbf{30}, 1477 (1998). 

\bibitem{goswami1} R. Goswami, P. S. Joshi, Phys. Rev. D. \textbf{76}, 084026 (2007), 

\bibitem{magli1} G. Magli, Class. Quant. Grav. \textbf{14}, 1937 (1997).

\bibitem{magli2} G. Magli, Class. Quant. Grav. \textbf{15}, 3215 (1998).

\bibitem{giambo1} R. Giambio, F. Giannoni, G. Magli and P. Piccione, Commun. Math. Phys. \textbf{235}, 563 (2003).

\bibitem{harada1} T. Harada, K. Nakao and H. Iguchi, Class. Quant. Grav. \textbf{16}, 2785 (1999).

\bibitem{harada2} T. Harada, H. Iguchi, K. Nakao, Prog. Theor. Phys. \textbf{107}, 449 (2002).

\bibitem{joshi7} P. S. Joshi, D. Malafarina	Int. J. Mod. Phys. D, \textbf{20}, 2641 (2011).

\bibitem{giambo} R. Giambo, Fabio Giannoni, Giulio Magli and Paolo Piccione, Gen. Relativ. Gravit. \textbf{36}-6, 1279 (2004). 

\bibitem{ori1} A. Ori and T. Piran, Phys. Rev. Lett. \textbf{59}, 2137 (1987).

\bibitem{ori2} A. Ori and T. Piran, Gen. Relativ. Gravit. \textbf{20}, 7 (1988).

\bibitem{ori3} A. Ori and T. Piran, Phys. Rev. D \textbf{42}, 1068 (1990).

\bibitem{carr} B. J. Carr, A. A. Coley, M. Goliath and U. S. Nilsson, C. Uggla, Class. Quant. Grav. \textbf{18}, 2 (2001)

\bibitem{joshi4} P. S. Joshi, D. Malafarina, Gen. Relativ. Gravit. \textbf{45}, 305 (2013).

\bibitem{harada} T. Harada, Phys. Rev. D \textbf{58}, 104015 (1998).

\bibitem{goswami} R. Goswami and P. S. Joshi, Class. Quantum Grav. \textbf{21}, 3645 (2004).

\bibitem{joshi5} P. S. Joshi, D. Malafarina, R. V. Saraykar, Int. J. Mod. Phys. D. \textbf{21}, 08, 1250066 (2012).
 
 
 \bibitem{dover} I. N. Sneddon, \textit{Elements of Partial Differential Equations}, (Dover Publications Inc., Mineola, New York, (2006). 
 
 \bibitem{bernstein} D. L. Bernstein, \textit{Existence Theorems in Partial Differential Equations.}, Annals of Mathematics Studies, no. 23, (Princeton, N.J., 1950).

\bibitem{lemaitre} G Lemaitre, Ann. Soc. Sci. Bruxelles I A \textbf{53}, 51 (1933).

\bibitem{tolman} R. C. Tolman, Proc. Natl. Acad. Sci. USA \textbf{20}, 410 (1934). 

\bibitem{bondi} H. Bondi, Mon. Not. Astron. Soc. \textbf{107}, 343 (1947).

\bibitem{joshi6}  P.S. Joshi, \textit{Global Aspects in Gravitation and Cosmology}, (Clendron Press,
Oxford, 1993).

 \end{thebibliography}
\end{document}